\documentclass[10pt,letterpaper]{article}
\usepackage[top=0.85in,left=2.75in,footskip=0.75in]{geometry}
\usepackage{amsmath,amssymb}

\usepackage{changepage}

\usepackage[utf8x]{inputenc}

\usepackage{textcomp,marvosym}

\usepackage{cite}

\usepackage{nameref}
\usepackage{ragged2e}

\usepackage{microtype}
\DisableLigatures[f]{encoding = *, family = * }

\usepackage[table]{xcolor}

\usepackage{array}

\newcolumntype{+}{!{\vrule width 2pt}}

\newlength\savedwidth

\raggedright
\usepackage[aboveskip=1pt,labelfont=bf,labelsep=period,justification=raggedright,singlelinecheck=off]{caption}

\makeatletter
\renewcommand{\@biblabel}[1]{\quad#1.}
\makeatother

\usepackage{lastpage,fancyhdr,graphicx}
\fancyheadoffset[L]{2.25in}
\fancyfootoffset[L]{2.25in}
\begin{document}
\vspace*{0.2in}

\begin{flushleft}
{\Large
\textbf\newline{Social and Child Care Provision in Kinship Networks: an Agent-Based Model} % Please use "sentence case" for title and headings (capitalize only the first word in a title (or heading), the first word in a subtitle (or subheading), and any proper nouns).
}
\newline
\\
Umberto Gostoli\textsuperscript{1\Yinyang},
Eric Silverman\textsuperscript{1*\Yinyang}
\\
\bigskip
\textbf{1} University of Glasgow, MRC/CSO Social and Public Health Sciences Unit, Glasgow, G2 3AX, UK
\\
\bigskip

% Insert additional author notes using the symbols described below. Insert symbol callouts after author names as necessary.
% 
% Remove or comment out the author notes below if they aren't used.
%
% Primary Equal Contribution Note
\Yinyang These authors contributed equally to this work.

%\textcurrency Current Address: Dept/Program/Center, Institution Name, City, State, Country % change symbol to "\textcurrency a" if more than one current address note
% \textcurrency b Insert second current address 
% \textcurrency c Insert third current address

% Group/Consortium Author Note
%\textpilcrow Membership list can be found in the Acknowledgments section.

% Use the asterisk to denote corresponding authorship and provide email address in note below.
* eric.silverman@glasgow.ac.uk

\end{flushleft}
\section*{Abstract}

\justifying

Providing for the needs of the vulnerable is a critical component of social and health policy-making. In particular, caring for children and for vulnerable older people is vital to the wellbeing of millions of families throughout the world. In most developed countries, this care is provided through both formal and informal means, and is therefore governed by complex policies that interact in non-obvious ways with other areas of policy-making.   
In this paper we present an agent-based model of social and child care provision in the UK, in which agents can provide informal care or pay for private care for their relatives.  Agents make care decisions based on numerous factors including their health status, employment, financial situation, and social and physical distance to those in need.  Simulation results show that the model can produce plausible patterns of care need and availability, and therefore can provide an important aid to this complex area of policy-making.  We conclude that the model's use of kinship networks for distributing care and the explicit modelling of interactions between social care and child care will enable policy-makers to develop more informed policy interventions in these critical areas.

% Please keep the Author Summary between 150 and 200 words
% Use first person. PLOS ONE authors please skip this step. 
% Author Summary not valid for PLOS ONE submissions.   
%\section*{Author summary}
%Lorem ipsum dolor sit amet, consectetur adipiscing elit. Curabitur eget porta erat. Morbi consectetur est vel gravida pretium. Suspendisse ut dui eu ante cursus gravida non sed sem. Nullam sapien tellus, commodo id velit id, eleifend volutpat quam. Phasellus mauris velit, dapibus finibus elementum vel, pulvinar non tellus. Nunc pellentesque pretium diam, quis maximus dolor faucibus id. Nunc convallis sodales ante, ut ullamcorper est egestas vitae. Nam sit amet enim ultrices, ultrices elit pulvinar, volutpat risus.

\bigskip

\textit{``The moral test of government is how it treats those who are in the dawn of life, the children; those who are in the twilight of life, the aged; and those in the shadows of life, the sick, the needy and the handicapped."}\\
{\textit{--- Hubert Humphrey Jr.}}

%\linenumbers

% Use "Eq" instead of "Equation" for equation citations.
\section*{Introduction}
One of the most critical, and the most testing, tasks of modern society is the provision of personal and medical care for people who, due to their age or health conditions, are in a particular state of vulnerability and frailty. In particular, every society must provide \emph{child care} for the care needs of their children, and \emph{social care} for adults who need help with their activities of daily living (ADLs).  In most developed countries, the state plays an important role in the provision of care for these vulnerable groups.  However, formal and informal care provided within the household or broader kinship network is often critical to the health outcomes of vulnerable people. As populations of older people continue to increase while birth-rates drop in developed countries, some governments are confronted by a substantial increase in the demand for care.

In the UK the supply of carers is decreasing over time as birth-rates drop, even while the increasing elderly population requires ever more support \cite{coleman2002replacement}. A recent Age UK report states that almost 50\% of over-75s are living with a long-term illness that limits their ADLs \cite{age2017briefing}.  Given that this age group is among the fastest-growing in the country, expectations are that the demand for care will outpace the available carer population.

Consequently, \emph{unmet care need} is of critical importance to health and social care policy-making in the UK.  Ipsos MORI reports that a majority of the aged with care needs have at least some unmet care needs \cite{coleman2002replacement}, while Age UK estimates that 1.2 million people received insufficient care in 2017 \cite{age2017briefing}.  Carers UK estimates that in order to meet the skyrocketing levels of care demand, the population of carers would need to increase by 40\% over the next 20 years \cite{carers2015facts}. According to Wittenberg and Hu (2015), demand for privately-funded social care is also expected to rise significantly over a similar period, with expenditure on private care to nearly triple by 2035 \cite{wittenberg2015projections}.

For the majority of households with social care needs, the problem of meeting these needs is compounded by the necessity of meeting their family's child care requirements. According to FullFact, 79\% of families in England with children aged 0 to 14 used some form of childcare, with 66\% of them using formal childcare, 40\% using informal childcare and 28\% using both \cite{childcareCost2017}. Further, according to the OECD report \emph{Society at a Glance 2016}, UK families spend over 30\% of their income on childcare \cite{oecd2016}.

The provision of social care in the UK is largely dependent on informal care, or care provided on a volunteer basis by family members. A 2018 report from the National Audit Office estimates the value of UK informal care at \pounds 100 billion per year \cite{national2018adult}. Aldridge and Huges, using data from The Family Resources Survey 2013/14, report that there were 5.3 million informal carers in the UK \cite{aldridge2016informal} and the Health Survey for England 2017 states that 68\% of participants aged 65 and over reported receiving help from unpaid helpers, while 21\% said they had received help from both unpaid helpers and paid helpers \cite{brown2018health}. In this regard, the importance of support and care-giving networks has long been recognized \cite{tennstedt1989informal, keating2003understanding}. Tennstedt et al. (1989) reported that informal care is provided mostly through networks of carers with an average of three to five members, predominantly composed of an individual's close relatives \cite{tennstedt1989informal}. 

Using data from the Family Resources Survey from 2011/12 to 2013/14, Aldridge and Huges find that 72\% of carers provide care to a member of their immediate family, i.e. a parent (40\%), partner (18\%), children (14\%) \cite{aldridge2016informal}. Similarly, Petrie and Kirkup estimated that around 51\% of carers provide care to a member of their own household \cite{petrie2018caring}. Using data from the Health and
Retirement Study 2011, Wettstein et al. show that 31\% of informal care in the US was provided by partners; 47\% by sons or daughters; and 18\% by other close relatives (e.g. children-in-law or grandchildren), with non-relatives contributing for just 4\% of the total informal care provided\cite{wettstein2017much}.

As for formal social care, the National Audit Office estimates that privately paid-for care amounts to approximately \pounds 11 billion in 2016-17, which increases to approximately \pounds 14 billions when we include private `top-ups' to the cost of the care arranged by local authorities. Empirical research has also shown that the type and amount of social care provided is affected by socioeconomic status. Petrie and Kirkup (2018) report that people working in routine occupations and those with lower qualifications are more likely to provide informal care \cite{petrie2018caring}.

Given the demographic trends outlined above, an increasing number of households will need to manage their resources to provide for both child care and social care needs, meaning that in these cases these two types of care are deeply interrelated.  In addition, both the social and child care provision processes taking place within these households, and their connected care-giving networks, are affected both directly and indirectly by the current government's child and social care policies.  With that in mind, we propose that understanding how child and social care need evolves over time, and the socioeconomic processes that underline the provision of care, are a vital component in any attempt to develop and implement effective and sustainable care policies.

In this paper, we present an agent-based model (ABM) of the UK informal and privately-funded formal care system, with the goal of capturing the complex relationships between social and child care, and the impact of social policies on these processes. This model provides a theoretical framework that enables us to improve our understanding of the complex care allocation system, where demographic, social and economic factors interact to determine the dynamics of care demand and supply. Further, using ABMs enables us to model scenarios of economic and social policy change, providing a means to test social policies which are meant to affect child and social care provision, and reveal any possible unintended side-effects (\emph{spillover effects}) of those policies prior to implementing them in the real world.

Previous work has explored social care provision and policy solutions using ABMs \cite{noble2012linked,silverman2013simulating}.  This simulation extends these efforts significantly, and models the provision of care not just as a simple transaction from one agent to another, but as a negotiation conducted across kinship networks with reference to numerous social, economic and geographical factors.  As a result, we propose that this model can support and inform child and social care policy-making more comprehensively than other methods.

\section*{Motivations}

Our primary motivation in this paper was not to generate point estimates of policy outcomes, but to develop a framework that could be capable of modelling the full complexity of social care.  At this stage our behavioural assumptions are subject to change, and will be further informed by policy-makers and practitioners in future iterations.  These results therefore should not be taken as policy advice, but instead as proof-of-concept work that demonstrates our model's potential to inform policy-making decisions relating to social care provision.  By documenting the model and its component processes in great detail in this paper and our previous work \cite{gostoli2019modelling}, we hope to inspire more agent-based modelling work in this area.      

Future work will incorporate more real-world data and insight from policy-makers, practitioners, and user groups, building upon the foundations described here.  Given the complexity of social care provision and the profound impact policy changes in this area have on the lives of families, we are proceeding methodically and cautiously in building and testing our framework before we begin using it to evaluate potential social care policy solutions.

\section*{Basics of the model}

In this section we provide a summary of the model's core economic and social processes.  This model is a comprehensive re-implementation and extension of previous work in Noble et al. \cite{noble2012linked} and Silverman et al. \cite{silverman2013simulating}, adding numerous processes and sub-processes to that basic framework.  Complete Python 2.7 source code for the simulation is available in our GitHub repository at \url{https://github.com/UmbertoGostoli/CareSim----Informal-and-Formal-Child-and-Social-Care/releases/tag/v0.8}. 

The modelling framework is under continuous development, and as such we recommend that any interested colleagues follow our updates on GitHub.  Releases will be produced periodically when new major features are added to the simulation.  
%\footnote{Complete Python 2.7 source code for the simulation is available here: \url{https://github.com/UmbertoGostoli/CareSim----Informal-and-Formal-Child-and-Social-Care/releases/tag/v0.8}}
Agents in the virtual UK depicted in this model occupy households, clusters of which form towns.  The sizes of these towns are set with rough correspondence to real UK population densities, scaled down by a factor of 1:10,000.  The simulation runs in one-year time steps; within each year processes taking place on a weekly scale are modelled.  The simulation begins in the year 1860, which allows sufficient time for the population dynamics to stabilise before 1951, at which point UK Census data is incorporated into the simulation. The simulation finishes in the year 2050.

Given the complexity of this simulation, we provide only brief summaries of some aspects which are explained in detail elsewhere, and refer readers to those papers for further information.  Changed and additional aspects of the current model are explained here in full.
%\footnote{This modelling framework is under continuous development, and as such we recommend any interested colleagues follow our updates on GitHub.  Releases will be produced periodically as new major features are added to the simulation.}.

\subsection*{Agent Life-Course}

Agents are classified as children (needing some form of child care) until the age of 11. At the age of 12 they become net providers of care and are classified as \emph{teenagers}. Agents enter adulthood at the working age of 16: at this point they can either start looking for work, or continue in education.  At the end of their education stage, agents become employed, with a salary which is a function of the socioeconomic status associated with the education level they have reached (see the Socioeconomic Status Groups subsection below). When agents reach the retirement age (set by a simulation parameter, with 65 as the default), they retire from employment and begin receiving a pension which is a fixed share of their final salary. If they retire earlier for health reasons, their pension is reduced accordingly.  Mortality rates in the model follow Noble et al. \cite{noble2012linked} and use a Gompertz-Makeham mortality model until 1951.  From that point we use mortality rates drawn from the Human Mortality Database \cite{hmd_2011}.  Lee-Carter projections generate agent mortality rates from 2009.

\subsection*{Partnership Formation and Dissolution}

Once they reach working age, agents can form partnerships.  Agents are paired randomly with probabilities that depend inversely on the agents' geographical distance from one another, their age and socioeconomic differences.  Model parameters set the relative weights of these factors.  Divorce probabilities are age-specific and are checked yearly to determine whether agents decide to divorce. Age-specific annual divorce probabilities determine whether a couple dissolves their partnership.  Fertility rates are computed similarly to mortality rates:  data from the Eurostat Statistics  Database \cite{esd_2011} and the Office for National Statistics \cite{ons_1998} are used from 1950--2009, with Lee-Carter projections taking over thereafter.

\subsection*{Internal migration}

Agents can migrate domestically for several different reasons (see the section Model Enhancements below).  Household relocations happen most frequently due to agents finding a partner or a new job in a different town.  Male agents will also relocate to new houses once a partnership dissolves, and any children produced by that partnership stay with the mother.  Retired agents with care needs may move in with one of their their adult children, with a probability determined by the their care need level and the amount of care supply in their child's household.  Orphaned children are adopted by a household in their kinship network, or by a random family if there are no available households in their kinship network.\\

\subsection*{Health status and care need}

Agents start their lives in a state of good health, and later may enter a state of care need according to gender- and age-specific probabilities.  The five categories of care need (which will be referred to as \emph{care need levels} in this paper) and the amount of hours per week of care required at each level are shown in Table~\ref{tab:careLevels}.  We assume that, once agents develop a health condition associated with a certain level of care need, they do not recover but progress to more severe conditions (and so, to higher levels of need) over time.  The chance of agents progressing to higher care need levels increases with age and with the sum of the agent's past unmet care needs (and decreases with higher socioeconomic status, see the \emph{Socioeconomic Status Groups} subsection below).  We thus assume that long periods of unmet care need will increase frailty, and that higher income and wealth allows for high-quality care to be purchased to increase quality-of-life.

\begin{table}[ht]
\centering
%\caption{The different care need categories, with the number of hours of care required per week\label{tab:careLevels}}
\caption{\bf Care need categories/levels and number of hours of care required.}
\begin{tabular}{ccc}
\hline
Care need category & Care need level & Weekly hours of care required \\ \hline
None & 0 & 0\\
Low & 1 & 8\\
Moderate & 2 & 16\\
Substantial & 3 & 36\\
Critical & 4 & 84\\
\hline
\end{tabular}
\label{tab:careLevels}
\end{table}

\section*{Model Enhancements}
The model we present in this paper is an offshoot of the Linked Lives model presented in Silverman et al. \cite{silverman2013simulating}, further extended in Gostoli and Silverman \cite{gostoli2019modelling} where the following features were introduced: socio-economic status (SES) groups; kinship networks; relocation's decision-making; formal (i.e. privately paid-for) care; public social care; a salary function; and hospitalization probabilities (which depend positively on levels of unmet care need). 
We provide very brief summaries of the 2019 additions of SES, kinship networks, the salary function, formal and public care provision aspects here, and refer the reader to Gostoli and Silverman \cite{gostoli2019modelling} for more details.  Subsequently we will describe the enhancements made to the current version in full. 

\subsection*{Socioeconomic Status Groups}
Agents are placed in one of five socioeconomic status groups (SES groups), based on the Approximated Social Grade from the Office for National Statistics.  These groups were redistributed as in Gostoli and Silverman \cite{gostoli2019modelling}. Each SES group is associated with an education level.  From the age of 16, an agent can decide whether to continue its studies or start searching for a job, in which case the agent is assigned the SES group associated with the education level he has reached. This choice is made by the agents every two years, until the age of 24 (i.e. at ages 16, 18, 20 and 22), with the probability of moving further up the education ladder depending on the household' income and the parents' level of education.  We assume that each education step lasts two years; each stage corresponds roughly to the UK education levels of A-level, Higher National Diploma, Degree and Higher Degree.
%\footnote{We assume each education step lasts 2 years. The educational stages correspond roughly to the UK education levels: A-level, Higher National Diploma, Degree and Higher Degree.}

The introduction of SES groups has a number of effects on the various stages of agent life-courses: a higher SES is associated with lower mortality and fertility rates; higher hourly salaries; lower salary growth rate. SES affects the agents' wealth, which is randomly assigned to agents according to their accumulated salaries (net of the expenses for social care) to reproduce the 2016 UK wealth distribution. The probability of two people to get married depends inversely on their `socioeconomic distance' (the other two factors being the geographical distance and the age difference). An agent's SES affects the agent's probability of transition to higher levels of care need. Moreover, the socioeconomic position of an agent affects its behaviour as care supplier (and that of the household it belongs to) through the agent's income, as we assume that the share of income allocated to care supply increases with the household's per capita income. %(see the Formal Care section below).

\subsection*{Kinship Networks}
Agents and households are associated with kinship networks, which are networks of households whose inhabitants have a consanguineous or affinal relationship with the agent or household the network is associated with.  We define `degrees' of kinship based on the network distance $D$ between households in the network; this kinship distance value ranges from 0 (same household) to III (uncles/aunts and nieces/nephews).

When an agent in a particular household is in a state of care need, the size of the kinship network associated with that agent, the kinship distances characterizing the kinship relations, and the individual states of the members of the households which are part of the agent's network determine the supply of care available to that agent. Table~\ref{tab:careAvailable} shows the hours of care supply associated with each agents' status and network distance:  

\begin{table}[ht]
%\begin{table}[!h]
\centering
\caption{
{\bf Amount of care agents can provide depending on their status and kinship  distance from the care receiver.}}
\begin{tabular}{ccccc}
\hline
Agent status & Household (D-0) & D-I & D-II & D-III \\ 
\hline
Teenager & 12 & 0 & 0 & 0\\
Student & 16 & 8 & 4 & 0\\
Employed & 16* & 12* & 8* & 4*\\
Retired & 56 & 28 & 16 & 8\\
\hline
\end{tabular}
\label{tab:careAvailable}
	%\begin{tablenotes}
	\\
      \small
        \textbf{*} Employed agents can provide additional care if they choose to reduce their working hours (i.e. in case it is more convenient than using income to pay for formal care. See the Formal Care section for details).
    %\end{tablenotes}
\end{table}

Physical distance also affects care provision, as we assume that only households in the same town as the care receiver can provide informal care.  In addition we assume that formal care is restricted by kinship distance, with provision of privately paid-for care occurring only among members of the same household or, if living in different households, only between parents and children.  

\subsection*{Relocation Decision-making}
Apart from care provision, kinship networks also influence the households' relocation decisions, as we assume that agents prefer to relocate to towns where more of their kinship network lives.  Each town is characterized by a \emph{total attraction}, one component of which is the town's \emph{social care attraction}, which is a growing function of the amount of care the household can expect to receive from (or supply to) the part of its kinship network living in that town. The other components determining a town's total attraction are: housing availability and cost (where the housing costs are represented by the Local Housing Allowance rates, with the rate being a function of the town's location and the household's size); and the town's SES profile (as we assume that agents prefer to relocate to towns with a relatively higher share of population of their own SES, or higher). Apart from the tows' attractions, the probability of relocating depends negatively on the \emph{relocation cost}, a measure of the social capital developed by the household's members in their current town which is a function of the number of household members and the number of years they have been living in their current town. The assumption underlying the relocation cost is that people develop valuable social capital in the town they live, which is largely lost when they relocate to other towns.

%We model household relocation as a two-stage process. In the first stage, the household samples a destination, with probabilities which are a function of the towns' total attraction. In the second stage, the household decides whether to relocate to this new town or to remain in its current location. The probability to relocate depends positively on the ratio between the destination's and the current town's total attractions, and negatively on the \emph{relocation cost}, which is a function of the number of household members and the number of years they have been living in their current town \footnote{The relocation cost can be conceptualised as a measure of the social capital developed by the household's members in their current town. This social capital would be lost by relocating to another town, so it acts as a barrier to relocation.}.

\subsection*{Salary function}
Every employed agent receives an hourly salary which is a function of its SES and its cumulative work experience, which is the discounted sum of all the shares of working hours during a week (i.e. if an agent always worked full time, this fraction is equal to 1). Formally, the salary function is the following Gompertz function:

\begin{equation}
w = Fe^{ce^{-rh}}
\end{equation}

where \textit{c = ln(I/F)}, \textit{I} is the initial hourly wage, \textit{F} is the maximum (or final) hourly wage, \textit{r} is the wage growth rate (with $I$, $F$ and $r$ being SES-specific parameters) and \textit{h} is the discounted cumulative work experience.
This salary function implies that if an agent takes time off work to provide informal care, this will result in less work experience and, therefore a lower hourly salary. On the other hand, given the properties of the care allocation mechanism, a lower hourly salary makes an agent more likely to provide informal care in the future, because of the lower value of its working time, compared to the working time values of other workers in the household and to the price of social care.

\subsection*{Government-funded social care}

Agents in a state of need may be entitled to publicly paid-for care, according to a government-funded social care scheme that mirrors the public social care scheme in force in England (for the sake of simplicity, at this stage we will not differentiate policies by region, although the spatially explicit framework we adopt makes this future development quite straightforward). On the basis of this scheme, all adults with a critical level of care need and whose level of savings is below \pounds 23,250 receive some public financial support. If their savings are below \pounds 14,250 the government pays all the social care expenses the care receiver cannot pay without reducing their income below \pounds 189 (called the \emph{minimum income guarantee}), whereas above this level of savings the amount paid by the government is reduced by \pounds 1 for every \pounds 250 of savings.  At this stage, our model does not distinguish between different forms of publicly paid-for formal care, i.e., between at-home care and care provided within care homes.
%\footnote{In this model, public financial support is used for any form of publicly paid-for formal care, without distinguishing the care received at home from that received in care homes.}

\subsection*{Formal Care}
Formal care is also allocated through the kinship network and can be bought using two financial sources: the care receivers' financial wealth and the households' income. For the latter, households allocate a share of their income to care for people living in the same household, or to first-degree relatives living in other households. We assume that the share of income allocated to care increases with the household's per-capita income %(i.e. households with a higher income allocate a higher share of it to provide for formal care).
The income allocated to care can be used either to buy privately paid-for care, or to take time off work to provide for informal care (in which case it represents not income spent but income not earned --  see the Care Allocation section below for details). The second source of privately paid-for care is the care-receiver's own financial wealth, which is a fixed share of  the agents' total wealth. We assume that the agents with care needs allocate a share of their financial wealth to formal care. The share allocated to formal care is positively related to the amount of financial wealth. As the agents buy formal care out of their financial wealth, it may eventually fall below the level at which the agents become entitled to government-funded social care.\\

\section*{Child Care}
In this updated version of the social care model from Gostoli and Silverman \cite{gostoli2019modelling}, we included another critical aspect of understanding care: child care provision (in this model, children are agents of age 0 to 11). In our model, we assume that all children, except newborns (agents of age 0), have the same child care need and that all children aged 1 to 11 have the same child care need, set to 56 hours per week. However, the \emph{net} care need of each child depends on his age, due to the presence of age-specific child care and education policies, which determine the quantity of child care provided by the state through nurseries and schools.
Newborns are treated as a special case, as they have a much higher need which is entirely supplied by their mother, who allocates all of her available supply of care to the newborn.

Although child care and social care seem similar on a surface level, there are deep differences between these two kinds of care which require us to treat them as two separate but interrelated processes. First, in the UK and most other developed countries there is a parental duty of care defined by law, while social care mostly rests on a social/moral obligation to care for one's relatives. Second, while child care is defined purely by the age of the recipient, social care implies a pathological condition which limits the recipients' activities of daily living.  Consequently, child care need is usually more predictable than social care need and can be supplied on a `one-to-many' basis, whereas social care usually is delivered on a `one-to-one basis'. Finally, due to this `one-to-one' characteristic of social care,  formal social care prices are between three and four times higher than formal child care.

These differences have important implications for the modelling of care provision. First, because of the legal frameworks related to the the provision of child care, we assume that it will have priority over the provision of social care, which therefore will be allocated the residual care supply remaining after the child care allocation process. Second, while social care need is linked to the single individuals needing care, we consider the child care need to be associated with the household rather than the individual children, and therefore characterise it by a certain amount of `aggregate' child care need whose structure depends on the number and age of the household's children. In other words, we assume that while social care is always personal, as it is provided directly to individuals, child care is provided to households.  

Finally, because of the different prices, we assume that, all else being equal, households will preferentially allocate their \emph{income} to provide for formal child care (i.e. the cheapest kind of care) and their \emph{time} to provide for informal care for the most expensive kind of care need. Although most of the time, the most expensive kind of care need will be social care need, because of the `one-to-many' nature of child care (that is, the possibility to satisfy multiple sources of child care need with each `time unit' of informal care), households with many children may find it more convenient to allocate their available time to provide for informal child care.  This in turn saves them the cost of multiple nursery fees, which may exceed the cost of formal social care depending on the number of children present. These differing characteristics mean that care provision is a complex social process, the computational implementation of which is discussed in the next section.

\section*{Care Allocation}
The care allocation we propose in this paper represents a complex negotiation conducted across kinship networks through which the two separated, but deeply interrelated processes of child and care provision take place.  Care allocation takes place in two stages: first the available care supply (composed of available time and income for care) is allocated to child care, and second the remaining resources are used to satisfy social care needs. 

In each stage the allocation process starts by randomly sampling a care-receiving unit (either a household with children, in the case of child care, or a person with care need, in case of social care) with a probability proportional to the unit's unmet care need. The care receiver is then associated with a care-giving household within the care receiver's kinship network (including the care-receiving household itself), sampled among all the potential care giving households with a probability proportional to the household's available care supply. This stochastic mechanism is based on the assumption that the higher the care need (care supply), the higher the probability of receiving (providing) care. 

%As for child care provision, we assume that, while informal care is provided through the household's kinship network, only the income of the children's own household is available to buy formal care. Therefore, the available care supply is given by the sum of informal and formal care supply for the children's household and just by the available informal care supply for all the other households in the kinship network (informal care supply which is zero if they live in other towns).
There are two main differences between social care and child care provision. First, while formal child care is provided only within the child's household, formal social care can be provided also through the income of households within distance 1 from the care recipient, in the care recipient's kinship network (i.e. the parents' and the children's households, if different from the care recipient's household). Second, besides the households' income, formal social care can be bought through another financial resource, which is the care recipient's own financial wealth. Therefore, the choice of care supply depends on the relative amounts of: a) time availability for all the households in the care receiver's kinship network living in the same town of the care receiver; b) the income of the households up to distance one from the care receiver; c) the care receiver's own financial wealth.

Once the care supplier has been selected, a 4-hour `quantum' of care is transferred from one member of the supplying household with available supply to the individual with care need (note that receiving and supplying agent may live in the same household). However, if a household within distance 1 from the care receiver is selected, a further decision needs to be taken about whether the time (i.e. informal care) or income (i.e. formal care) of the house is to be used.

While the selection of time will result in 4 hours of informal care being provided (and, correspondingly, 4 hours of time being subtracted from the care-giving household's time resources), if the resource selected is income, the care-giving household must decide whether to use income to buy formal care or to use the working time of a household's worker to provide for informal care (i.e. taking time off work). The choice depends on the hourly wage of the household's worker with the lowest wage: if it is lower than the price of formal care, the worker will prefer to take time off to provide care, whereas in the opposite case purchasing formal care is preferred.

Note that the way the prices of care are computed will differ between child and social care, due to the aforementioned `one-to-many' aspect of child care.  While the price of social care is fixed, the child care price to which the workers' wages are compared, in order to make the informal/formal care choice, is the price of formal child care multiplied by the number of children, because of the ability of the informal carer to satisfy multiple child care needs concurrently. We call these values \emph{informal child care values} (ICVs), representing the cost that the household avoids by providing informal child care. The underlying assumption is that while multiple children will increase the total cost of formal child care, informal care can satisfy multiple children's care needs per time unit provided, therefore allowing the household to avoid this cost. A household's ICVs depend on the household's number and ages of children and determine whether that household will elect to pay for formal care or take time off for informal care. 

When a household has both child and social care need, it will preferentially allocate informal care to the most expensive variety and formal care to the least (again, given the possibility of satisfying child care needs concurrently for multiple children, the relevant cost of child care in this regard are the household's ICVs). 
%Therefore, for the care allocation process, the household will separate its child care needs into two groups: the group associated with ICVs higher than the social care hourly rate; and the one associated with ICVs lower than the social care rate.  The high-ICV child care will be satisfied with informal care as much as possible; when the available supply is exhausted, the remainder will be satisfied with formal care when the ICV is higher than the supplier's hourly wage, and with informal care (provided by taking time off work) otherwise. Once the high-ICV child care need is satisfied, the low-ICV child care will be preferentially satisfied with formal care, unless enough informal supply (in relation to the household's social care need) exists to cover the need.  
After all the household's child care needs are satisfied, the remaining availability of time and income for care within the household's kinship network will be used to satisfy the household's social care need.

\section*{Social Policy Experiments}
Given the importance of child and social care provision to many families, most developed nations design and implement social policies intended to reduce the care burden on families and, in general, facilitate care provision. Furthermore, child care provision is affected by the education policies in place, to the extent that they affect the hours children spend in school.  In this model, we included the current child care, education and social care policies in force in England (neglecting, at this stage, differences between the UK regions). The inclusion of these policies and the related policy levers allows us to simulate care outcomes and costs under alternative social care policies, represented by different combinations of policy parameters, making this model a unique tool for developing and evaluating care policy interventions.

In these early-stage results, we investigated the effects of four policy interventions related to some key policy `levers' where policy-makers attempt to influence social care outcomes.  We developed four potential policy interventions designed to reduce the overall social care burden to UK society.  While these scenarios deal only with interventions affecting a single type of care at a time, child and adult social care are interrelated processes, so any policy targeting one type will affect the other. The four policy levers targeted by out policy intervention experiments (and their current values) are listed below:

\begin{itemize}
    \item Public child care cost contribution($\alpha$): the government adds an extra \pounds 2 for every \pounds 8 that working families spend on child care, up to £2,000 per child per year (i.e. there is a 20\% government contribution to child care costs).
    \item Hours of free child care ($\beta$): working families can get up to 1,140 hours per year of free child care for every child aged 3 and 4.
    \item Minimum care need for government-funded care ($\gamma$): local authorities pay the full social care cost of people with a critical level of social care need with savings of less than \pounds 14,250. If their savings are between this lower bound and \pounds 23,250, the person receiving the social care will contribute a pound for every \pounds 250 of savings to the weekly cost.
    \item Public social care cost contribution ($\theta$): the government contributes some fraction of the cost of social care.  Currently there is no such scheme in the UK.
\end{itemize}

In Table~\ref{tab:policyValues} we show the benchmark (default) and intervention levels of these four key parameters. In the Results section below, we compare the benchmark scenario with four policy scenarios, one for each intervention (leaving, in each scenario, the other policy levers at their benchmark level). In the first scenario, we increase the public child care cost contribution from 20\% to 80\% of the cost (i.e. the state refunds \pounds 8 for every \pounds 10 spent on child care); in the second scenario, the hours of free child care for children aged 3 and 4 are increased from 20 to 32 per week; in the third scenario, the minimum care need level for eligibility for publicly-funded social care is lowered from 4 to 3 (see Table~\ref{tab:careLevels}); in the fourth scenario, a public social care cost contribution scheme is introduced in which the state pays 50\% of the cost of social care.

\begin{table}[ht]
% \caption{Benchmark and intervention levels for the four policy levers\label{tab:policyValues}}
\centering
\caption{\bf Benchmark and intervention levels for the four policy levers}
\begin{tabular}{ccc}
\hline
Policy Lever & Benchmark & Intervention \\ \hline
alpha & 0.2 & 0.8 (P1)\\
beta & 20 & 32 (P2)\\
gamma & 4 & 3 (P3)\\
theta & 0 & 0.5 (P4)\\
\hline
\end{tabular}
\label{tab:policyValues}
\end{table}

We assume that the four policies are implemented from simulation year 2020 and compare the outputs of these four policy scenarios with the benchmark no-change scenario over the period 2020--2050. 

\section*{Results}
\begin{figure*}[!ht]
\caption{{\bf Population and tax payers.}}
\includegraphics[width=0.9\linewidth]{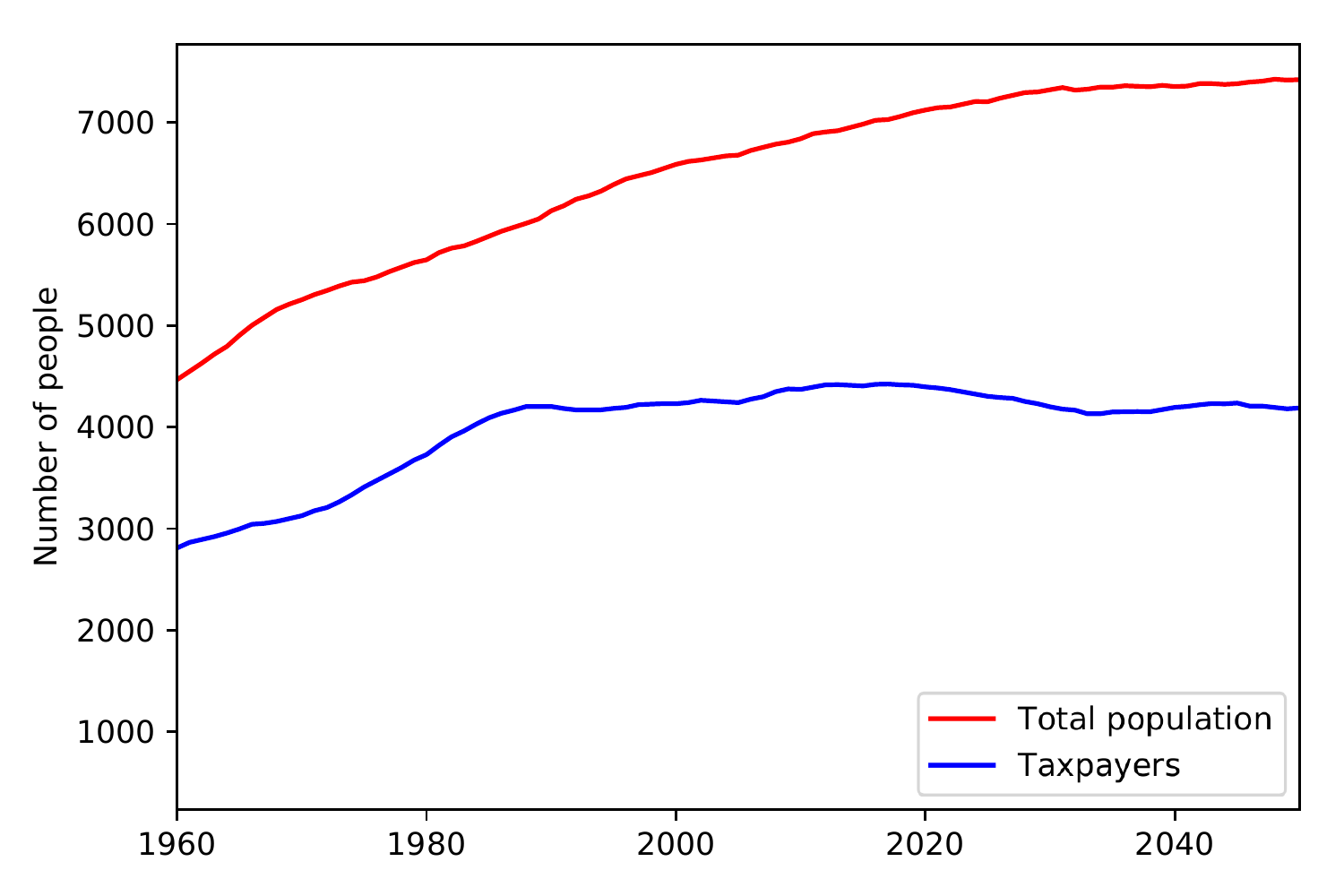}
\label{Fig1}
\end{figure*}

Here we present the outcomes of a representative `benchmark' simulation and compare some of these these to the effects of possible social policy interventions.
Figure~\ref{Fig1} shows the population and the proportion of tax-paying agents. Although the population keeps growing from 1960 to 2050, it grows at a decreasing rate. We can see that, after around 2030, the population growth curve becomes almost flat.
One of the main effects of population ageing can be seen from the dynamics of the working-age population, which grows even more slowly than the general population, as we can see from the growing gap between the latter (red line) and the former (blue line), starting from around 1990. 

\begin{figure*}[!ht]
\caption{{\bf Employment rate.}}
\includegraphics[width=0.9\linewidth]{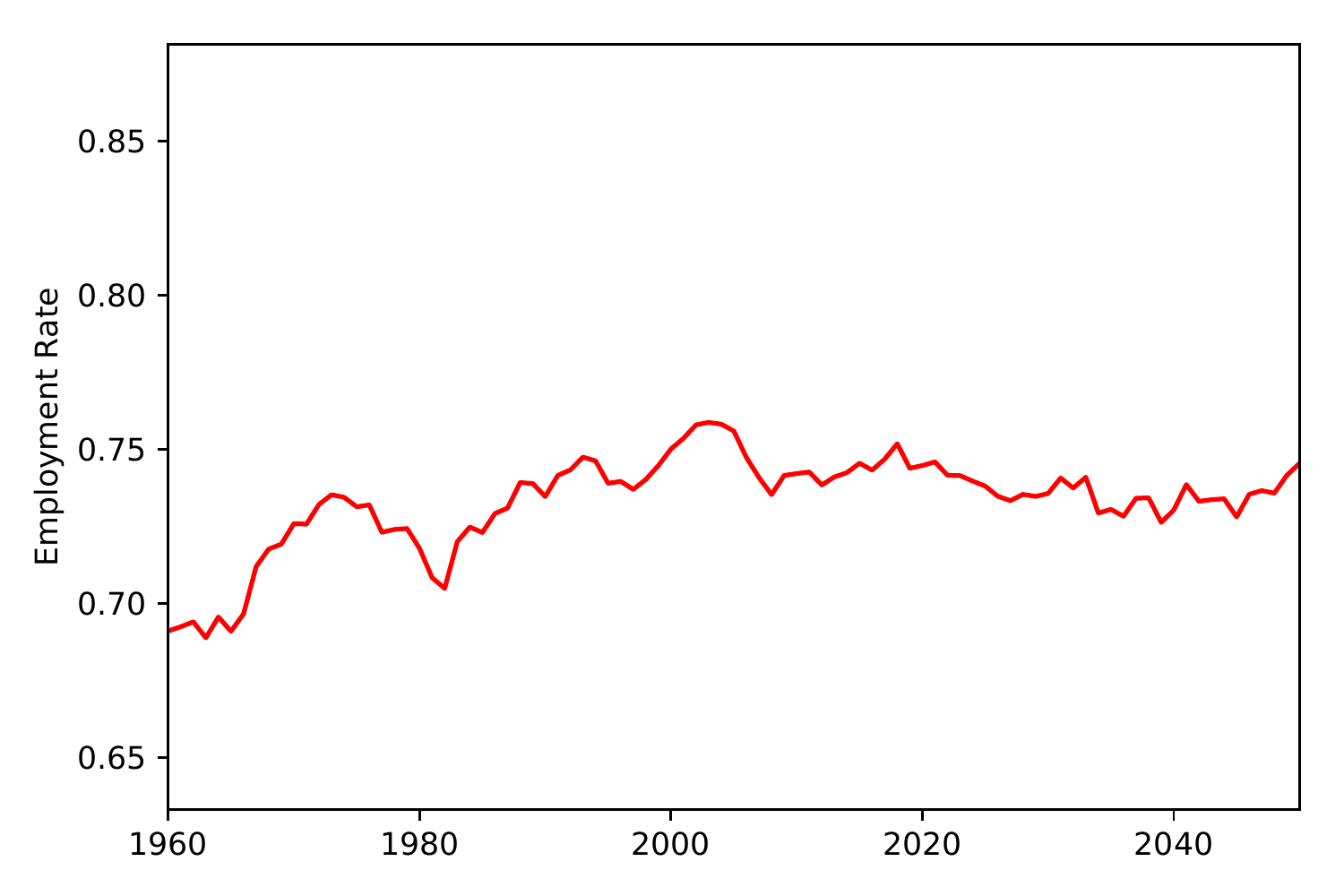}
\label{Fig2}
\end{figure*}

Figure~\ref{Fig2} shows the share of adult population (i.e. people from age 16 to age 65) who are employed (the others being students or having health conditions preventing them from working). We can see that it fluctuates mostly between 70 and 75\%, a level which is quite consistent with the empirical data.

\begin{figure*}[!ht]
\caption{{\bf Informal, formal, public care and unmet care need.}}
\includegraphics[width=0.9\linewidth]{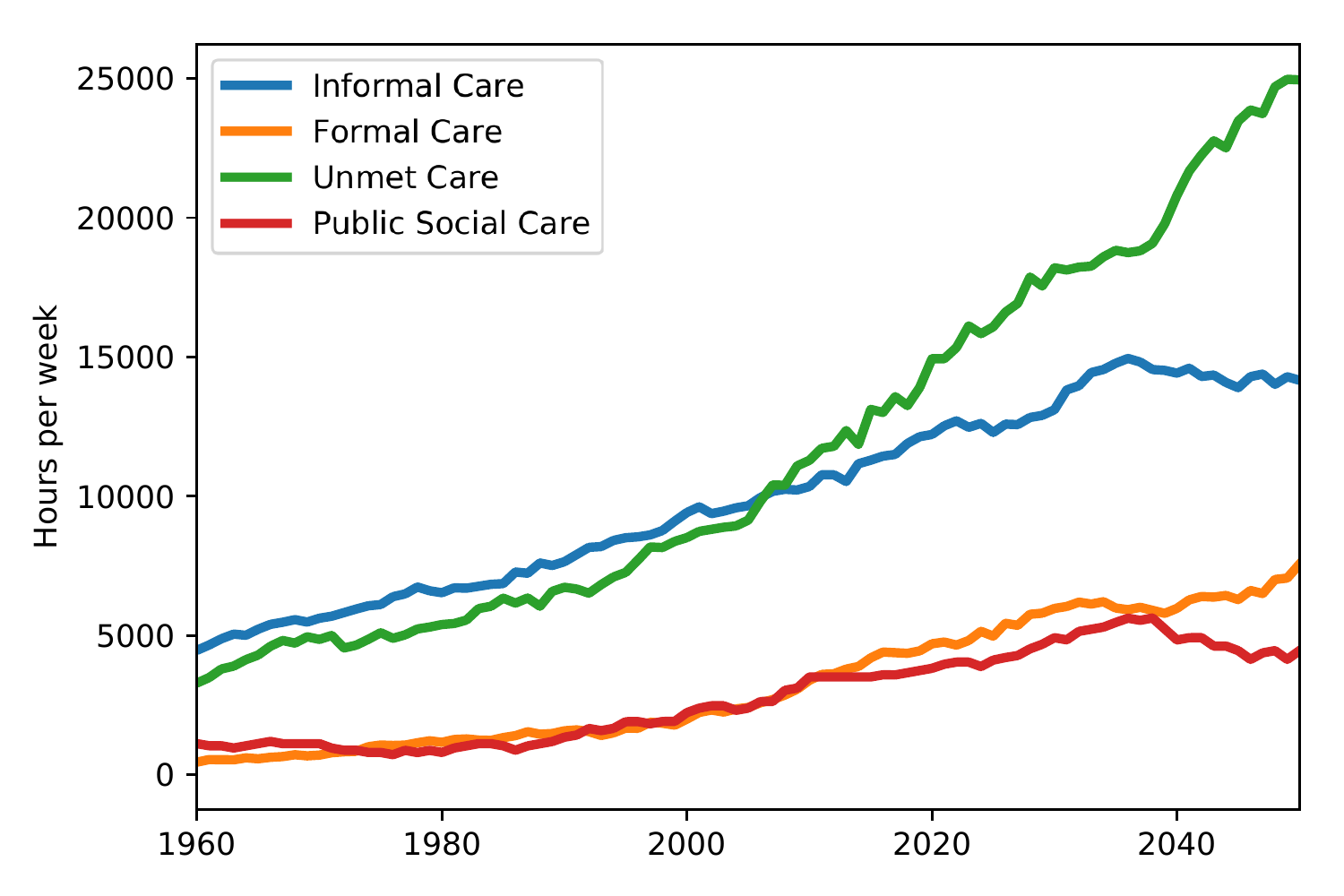}
\label{Fig3}
\end{figure*}

Figure~\ref{Fig3} shows the dynamics of the three kinds of social care supply we considered (i.e. informal, privately paid-for and public) and unmet social care need. We can see that although until the second half of the 2030s all the three kinds of care supply increase, supply cannot keep pace with the social care demand, as shown by the dynamics of the unmet care need. Moreover, our simulations show that in the second half of the 2030s the informal and public care supply will start to decrease (with the current social policies in place).

\begin{figure*}[!ht]
\caption{{\bf Total social care need.}}
\includegraphics[width=0.9\linewidth]{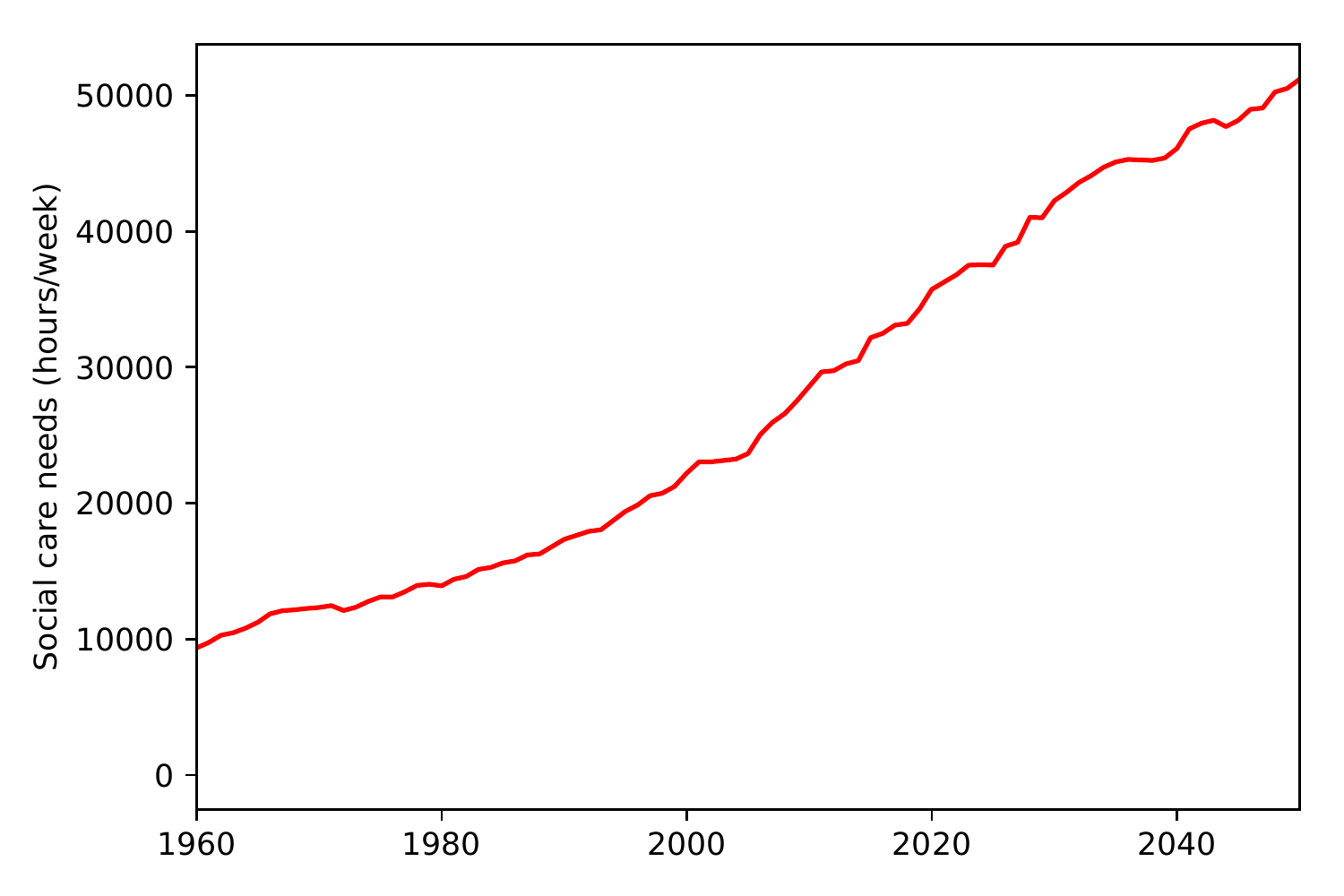}
\label{Fig4}
\end{figure*}

We can see more clearly the social care effects of these demographic trends in Figure~\ref{Fig4}, which shows the relentless and steady growth of social care need, with a slight increase in the growth rate around the year 2000. Our simulations show that social care need increased by a factor of 5 between 1960 and 2050, while the population increased just by a 1.6 factor in the same period. 

\begin{figure*}[!ht]
\caption{{\bf Total unmet social care need.}}
\includegraphics[width=0.9\linewidth]{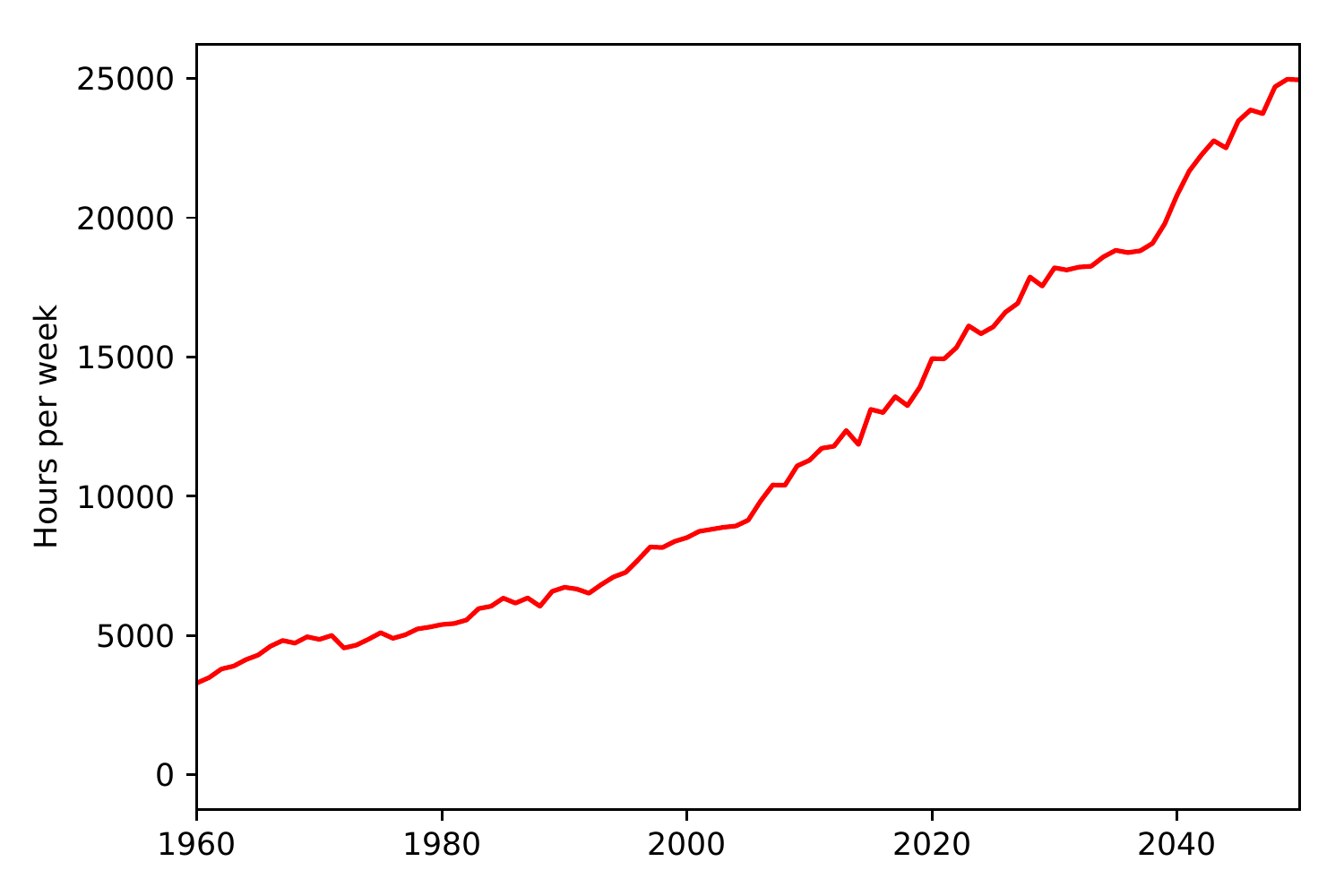}
\label{Fig5}
\end{figure*}

The dynamics of the unmet care need is shown more clearly in Figure~\ref{Fig5}, where we can see three stages characterized by increasing growth rates: the first stage ending around the year 2000; the second stage ending in the second half of the 2030s; a third stage, with the highest growth rate, from the second half of the 2030s to 2050. 

\begin{figure*}[!ht]
\caption{{\bf Average social care burden.}}
\includegraphics[width=0.9\linewidth]{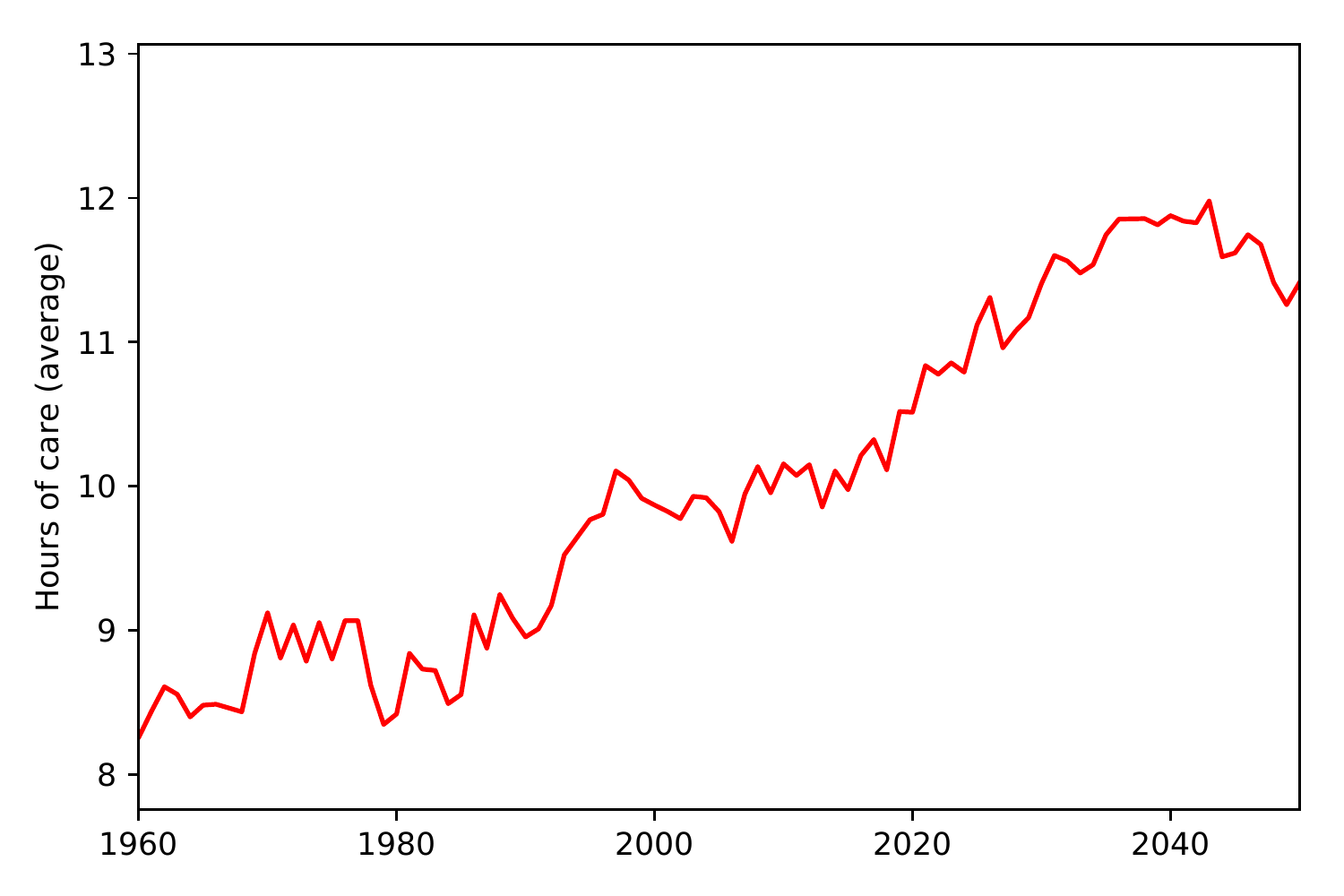}
\label{Fig6}
\end{figure*}

In Figure~\ref{Fig6} we can see that the increasing social care need causes the per capita hours of care delivered to increase, from just above 8 hours in 1960 to around 12 hours in 2040. After 2040, the average hours of care delivered decreases, a trend which reflects the dynamics of the informal care supply shown in Figure~\ref{Fig3}. Our simulations show that at the end of the 2030s, the demographic structure of UK society will be such that the number of people available to provide informal social care will decrease drastically.

\begin{figure*}[!ht]
\caption{{\bf Cost of public social care.}}
\includegraphics[width=0.9\linewidth]{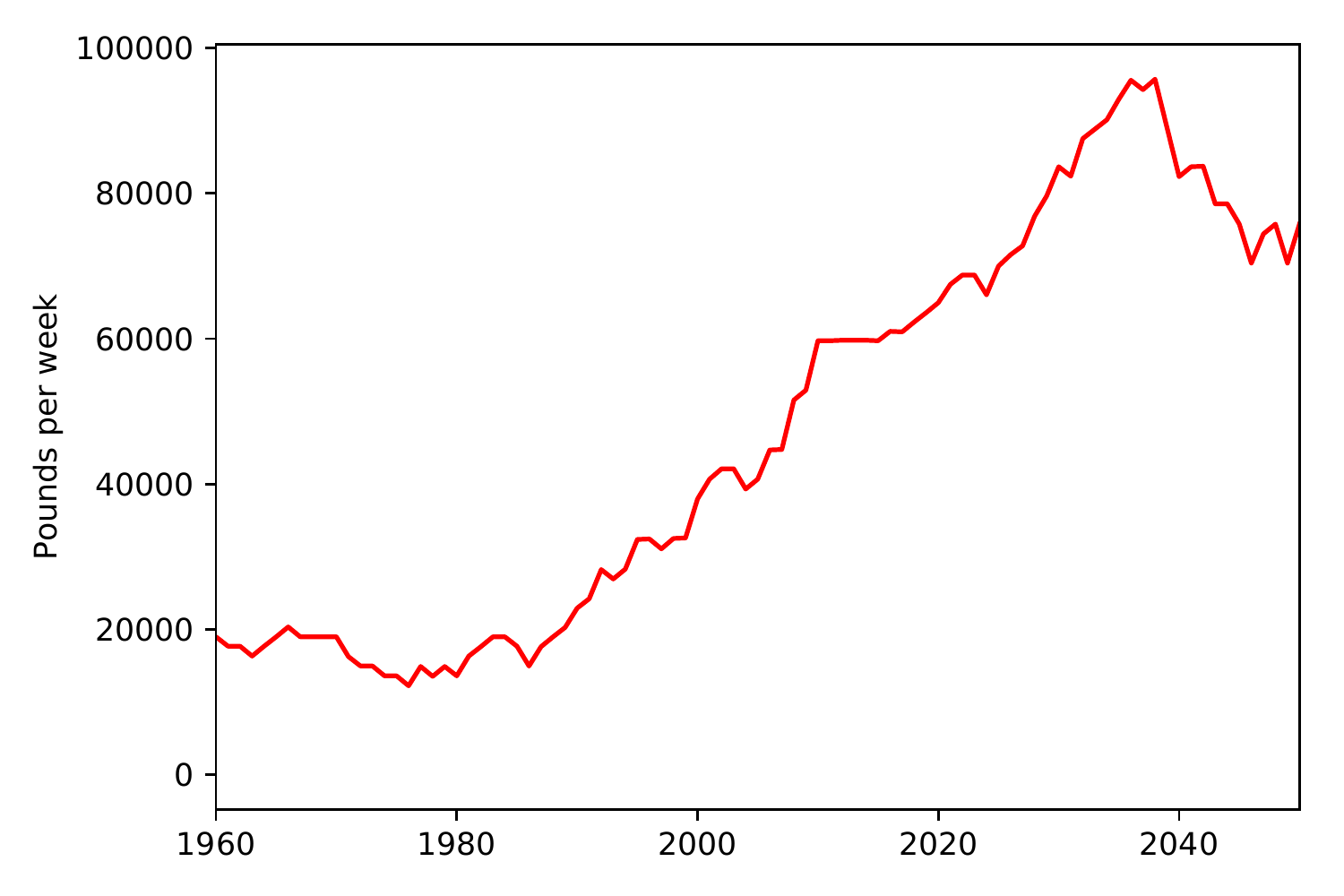}
\label{Fig7}
\end{figure*}

Figure~\ref{Fig7} shows the dynamics of the cost of public social care. We can see that it is composed of three phases: a steady phase up to the end of the 1980s; then a phase lasting until the end of the 1930s, in which the cost for public care grows by a factor of 5 with a relatively constant growth rate; a third phase were we can expect the growing trend of this cost to stop or even reverse due to changes in the demographic and socioeconomic structures.

\begin{figure*}[!ht]
\caption{{\bf Hospitalization costs.}}
\includegraphics[width=0.9\linewidth]{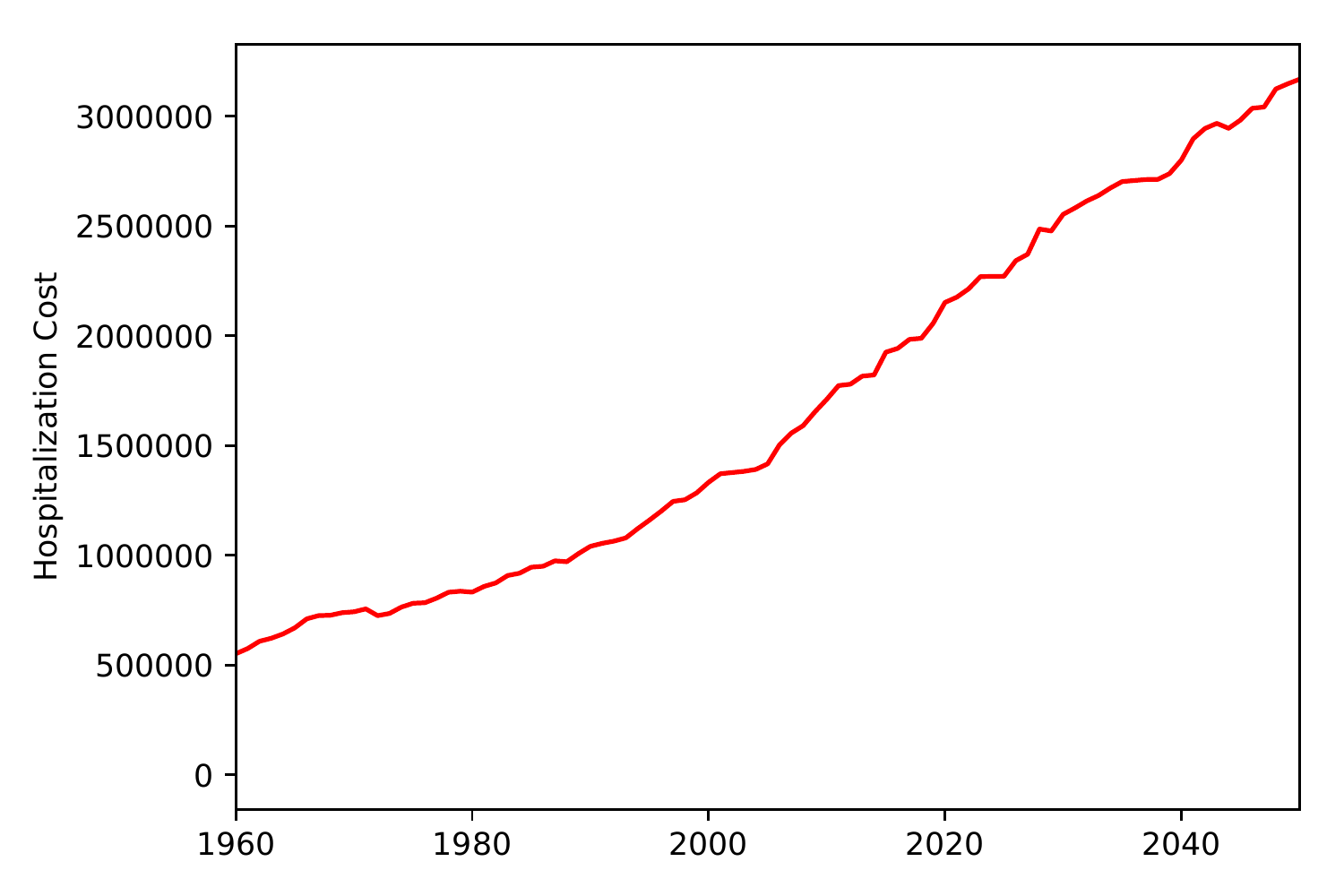}
\label{Fig8}
\end{figure*}

Differently from the cost of public social care, the cost of hospitalization increases with a relatively steady rate for all periods considered, as shown in Figure~\ref{Fig8}. From this figure we can see that, from 1960 to 2050, the hospitalization cost is expected to increase by a factor of 6.

\begin{figure*}[!ht]
\caption{{\bf Share of informal social care.}}
\includegraphics[width=0.9\linewidth]{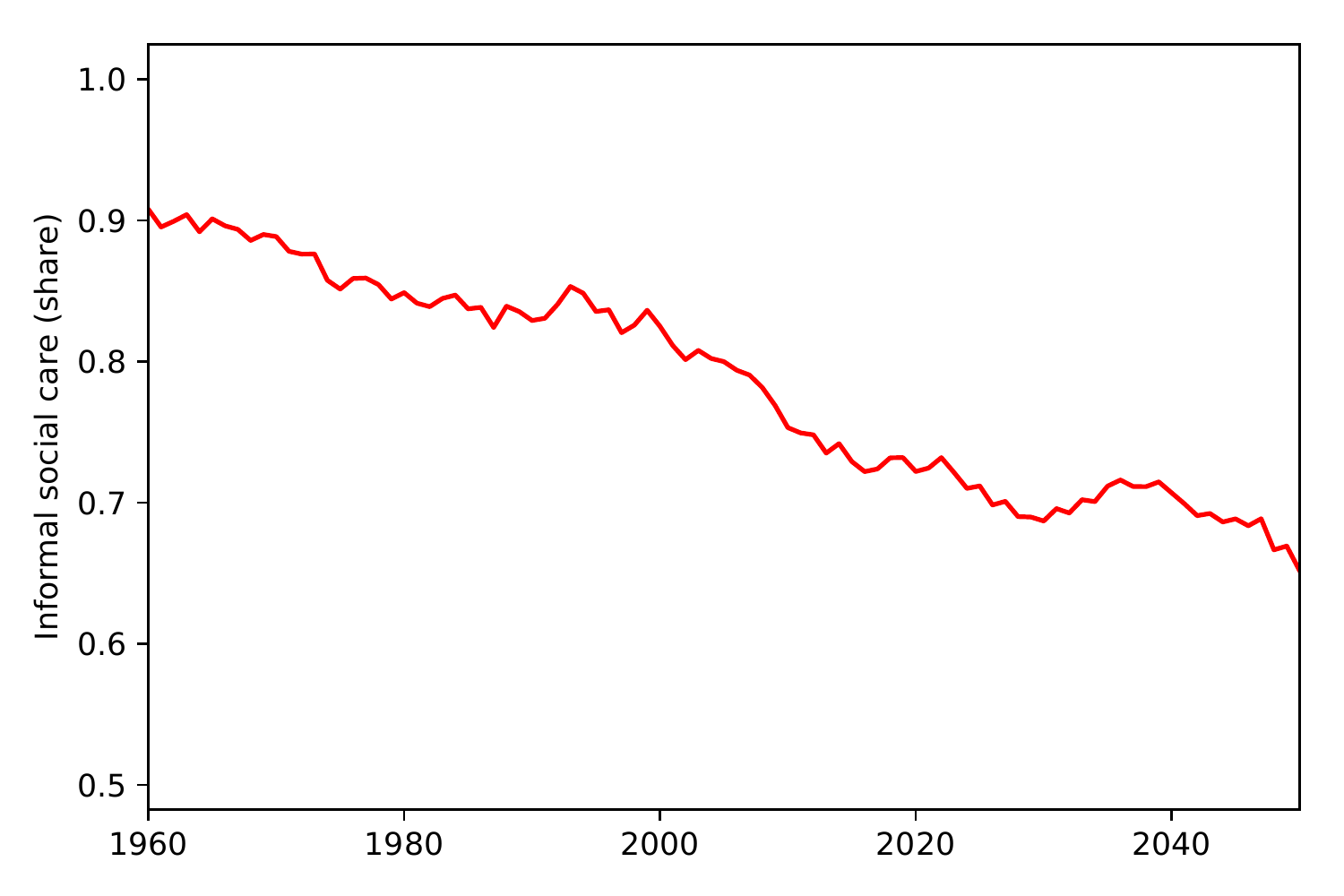}
\label{Fig9}
\end{figure*}

Figure~\ref{Fig9} shows that, due to changes to the demographic structure, the share of informal care supply over the total care supply is expected to decrease from 90\% in the 1960 to around 65\% in 2050. This means that the other forms of care supply grow at a faster rate than the growth rate of informal care, although not enough to satisfy the growing social care need. 

\begin{figure*}[!ht]
\caption{{\bf Gender pay gap.}}
\includegraphics[width=0.9\linewidth]{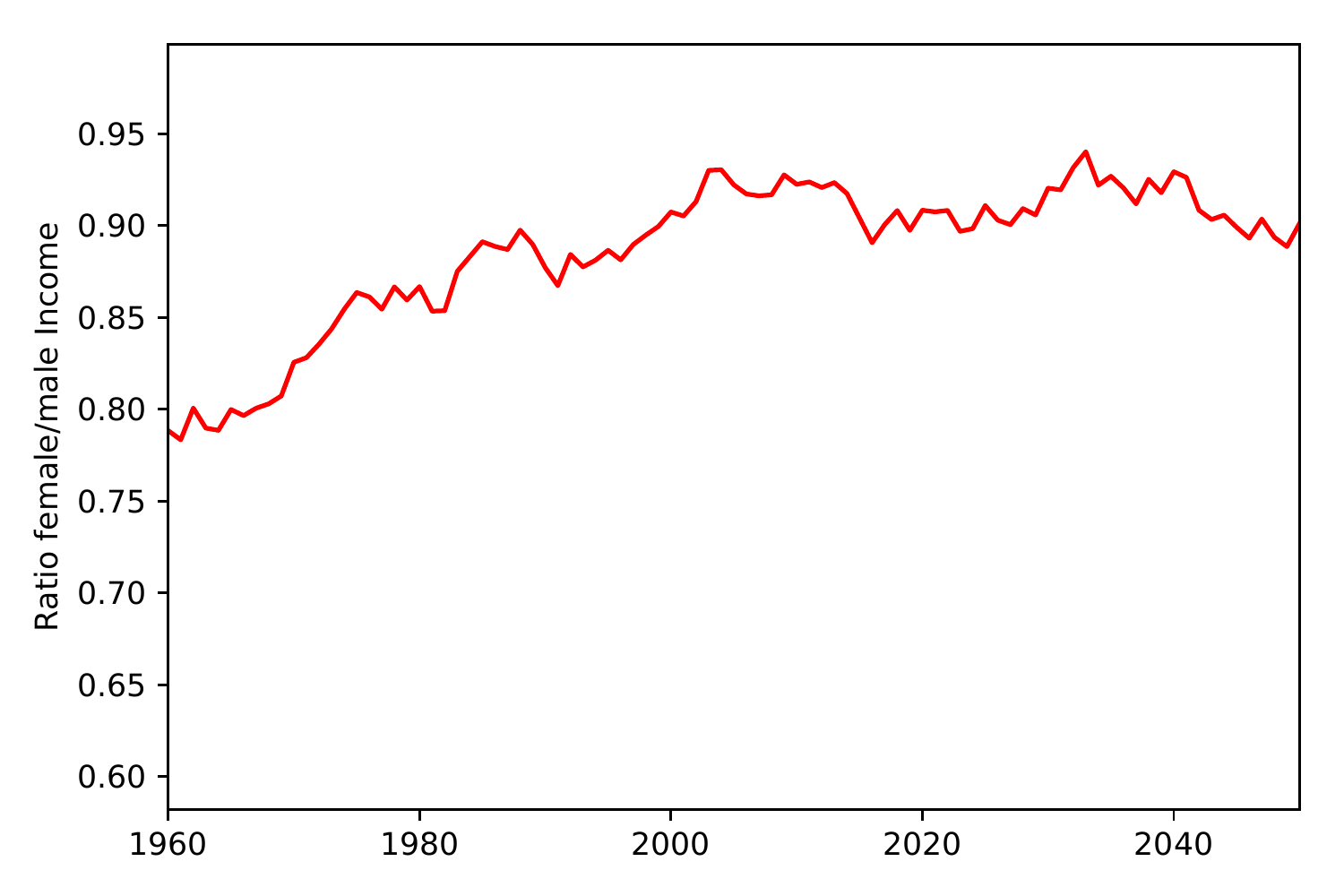}
\label{Fig10}
\end{figure*}

Finally, in Figure~\ref{Fig10} we show the dynamics of the gender pay gap, expressed as the ratio between female and male incomes. We can see that, in general, the simulation's outcome is quite consistent with the empirical data, showing a gender pay gap that fluctuates around 90\% after the year 2000 (meaning that pay for women is 90\% that of men).

\begin{figure*}[!ht]
\caption{{\bf Unmet care need: total period 2020-2050.}}
\includegraphics[width=0.9\linewidth]{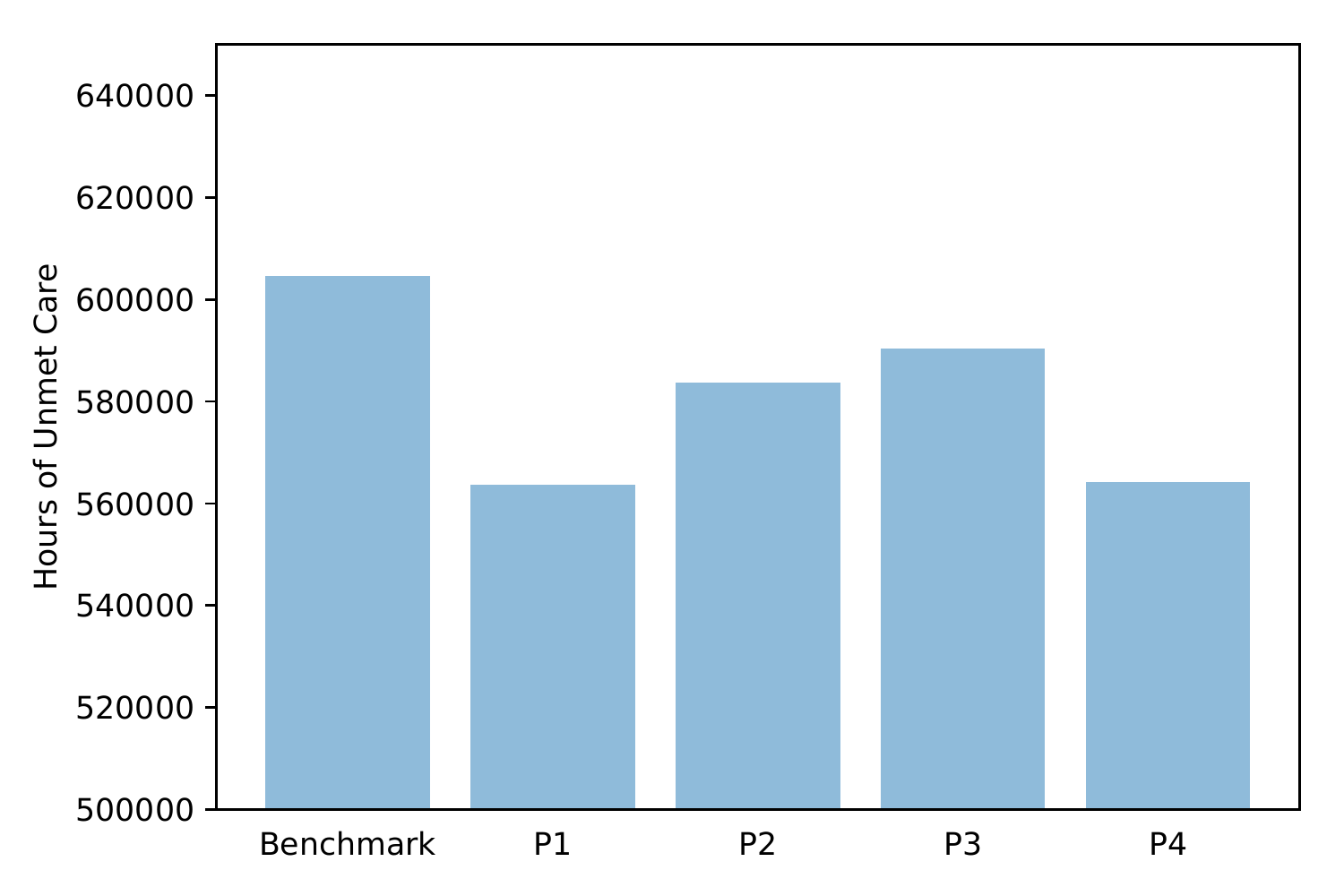}
\label{Fig11}
\end{figure*}

In the next ten figures, we will show the results of the social policy experiments, comparing the policy change outcomes with those of the \emph{current-policy} benchmark scenario. Figure~\ref{Fig11} shows the effects of the four policies, representing the parameters' change shown in Table~\ref{tab:policyValues}, on the total unmet social care need in the period 2020-2050. We can see that the first child care policy (i.e. the 80\% state contribution to child care cost, thereafter P1) is as effective as the last social care policy (i.e. the 50\% contribution to social care formal cost, thereafter P4) in reducing unmet care need. The positive effect of the child care policy change on the amount of unmet social care need is an example of a \emph{spillover effect}: a policy meant to affect primarily child care provision produces its effects on social care provision.  There are two main causes of this effect.  First, as child care becomes cheaper under intervention P1, for some families it will become more convenient to pay for formal child care rather than providing informal child care, so that more time resources will be available for informal social care.  Second, increased financial resources become available for formal social care, due to the reduced cost of formal child care. From this figure, we can also see that the increase in the hours of free child care (thereafter P2) and the inclusion of people with substantial social care needs among those eligible for public social care (thereafter P3) have a positive, although smaller, effect on unmet social care need. 

\begin{figure*}[!ht]
\caption{{\bf Policies' direct cost: total period 2020-2050.}}
\includegraphics[width=0.9\linewidth]{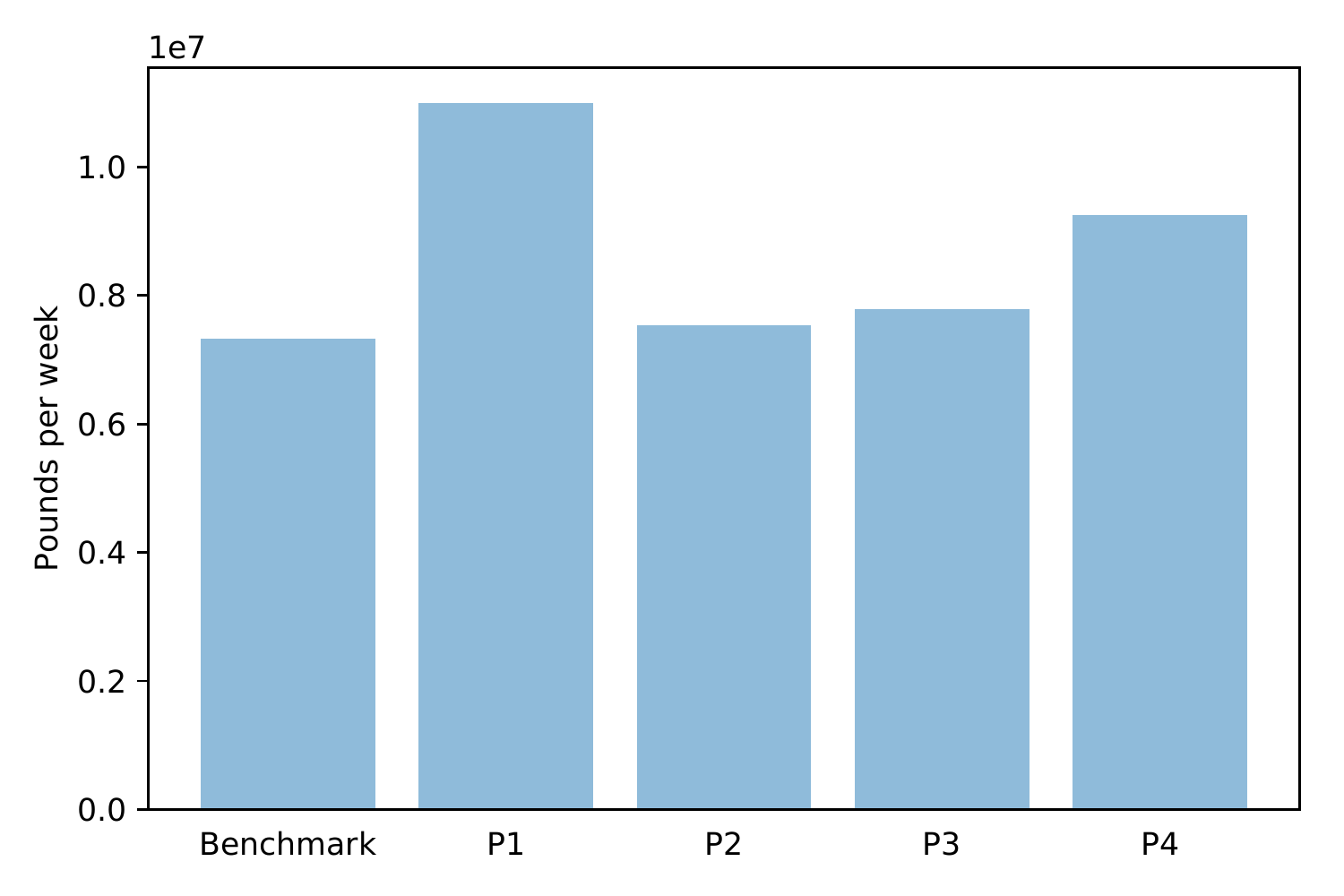}
\label{Fig12}
\end{figure*}

Figure~\ref{Fig12} shows the sum of the policy costs over the period 2020-2050. We can see from this figure that P1 is the most expensive policy, followed by P4. P2 and P3, on the other hand, have costs that are quite close to the benchmark scenario.

\begin{figure*}[!ht]
\caption{{\bf Social care need: total period 2020-2050.}}
\includegraphics[width=0.9\linewidth]{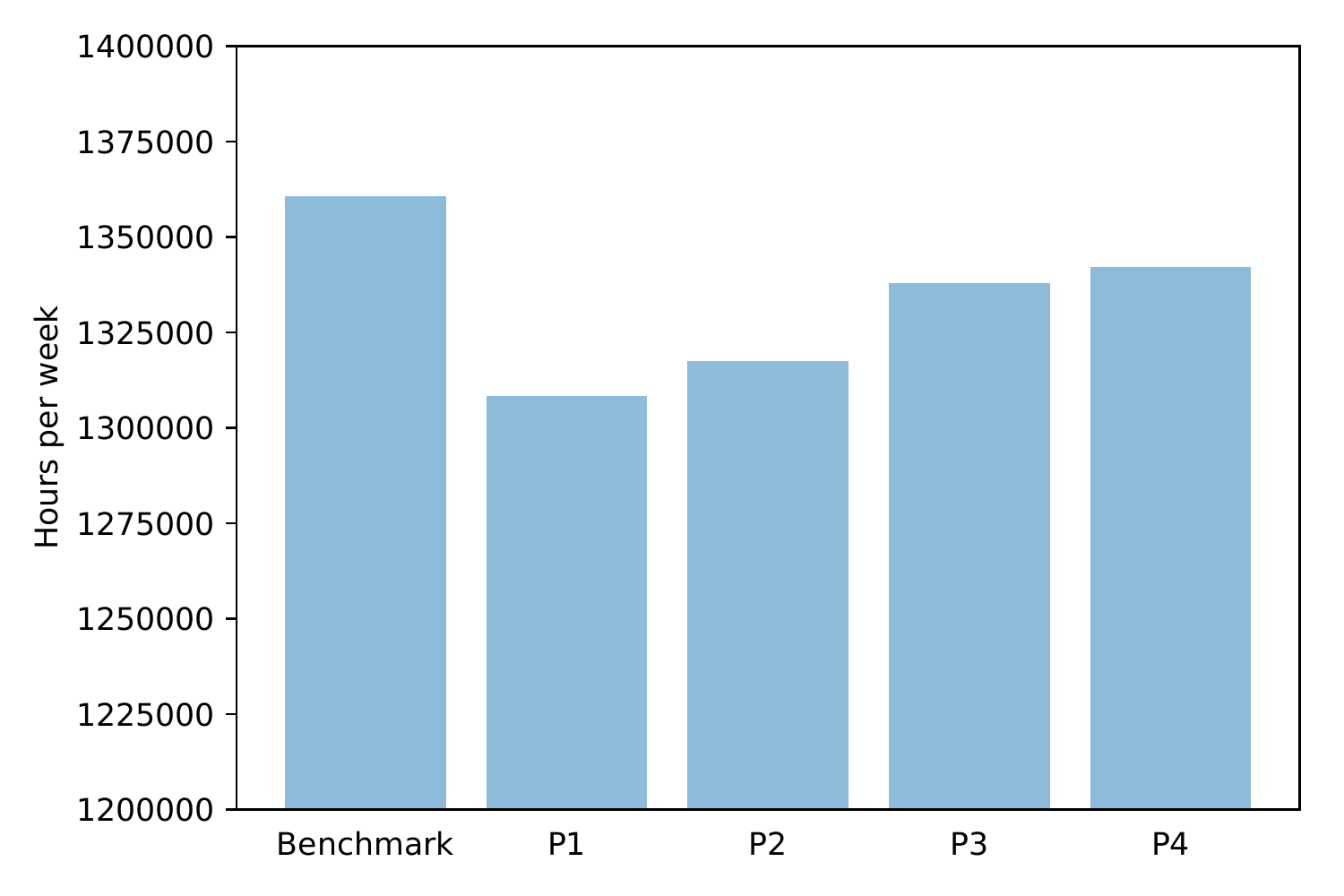}
\label{Fig13}
\end{figure*}

Figure~\ref{Fig13} shows the effects of the four policies on the total social care need in the period 2020-2050. From this figure, we can get some clues about the different ways the four policies (and especially P1 and P4) reduce the unmet care need. We can see that the child care policies reduce social care need more than the policies actually targeted at social care policies. So, our simulations show that  child care policies (in particular P1) decrease the unmet care need not by increasing the resources available for social care supply, but rather by reducing the social care demand. On the other hand, the social care policies (in particular P4), to the extent that they reduce the unmet care need, produce this effect mostly by increasing the care supply. The reduction of social care need, in turn, is an effect of the reduction of the unmet care need, as in the model the probability to develop more serious health conditions is positively related to the amount of unmet care need. 

The different effect of child and social care policies is due to the fact that while the considered social care policies tend to benefit only individuals with social care needs, and mostly those with the highest level of social care need, the child care policies benefit potentially all families, so that the additional resources are more likely to be used to provide social care to people who are, on average, at a lower level of care need.  A reduction of the unmet care need of people at lower levels of need reduces the probability that they will develop more serious conditions, and therefore reduces the total social care need; in contrast, an intervention aimed at people with the highest level of care need cannot allow for this reduction as the individuals concerned are already at the highest level of need (note that in our model we assume that individuals cannot improve in care need status, an assumption which reflects the fact that individuals in need of care tend to decline over time, due to the impact of complex conditions and increasing frailty). 

\begin{figure*}[!ht]
\caption{{\bf Hospitalization cost: total period 2020-2050.}}
\includegraphics[width=0.9\linewidth]{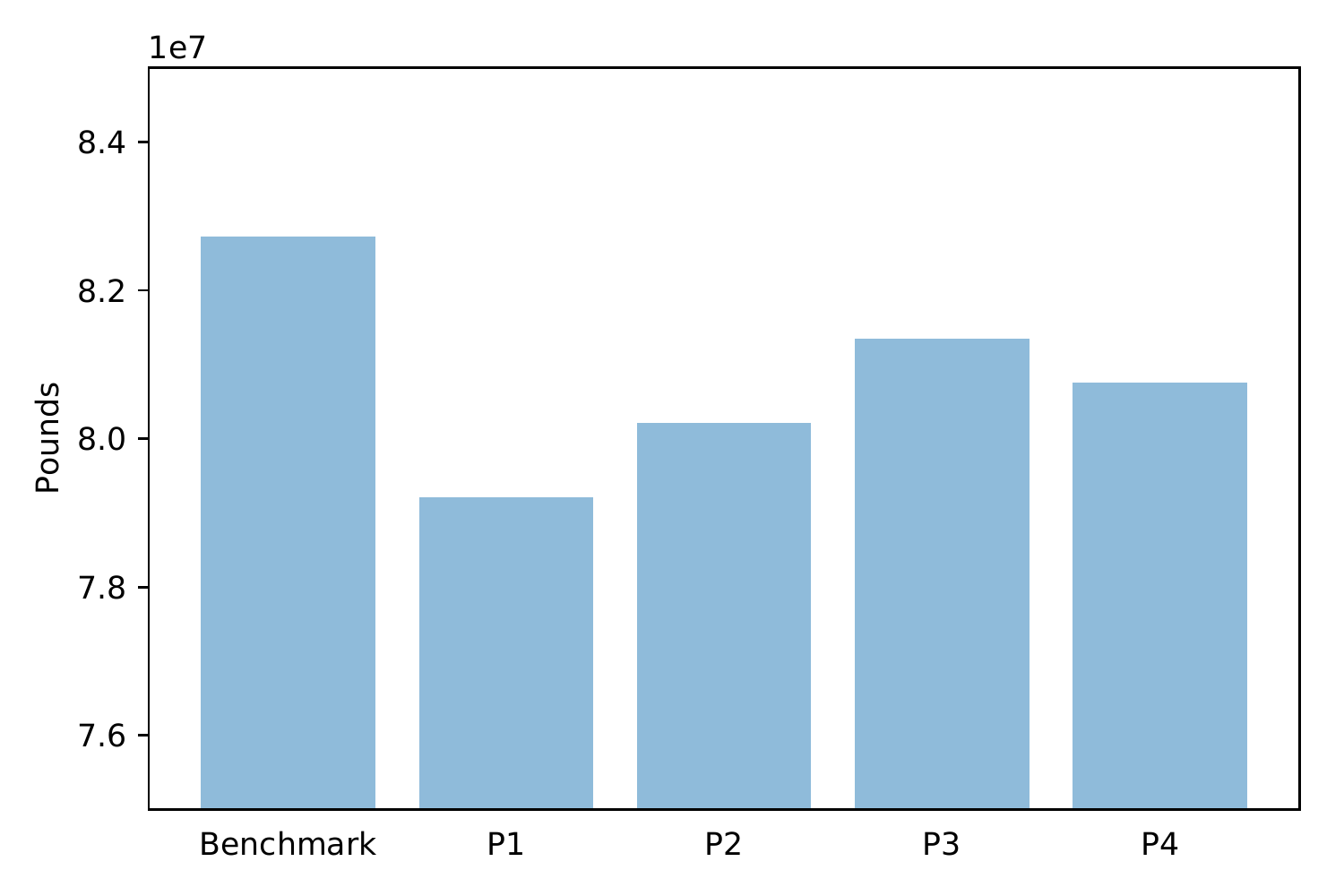}
\label{Fig14}
\end{figure*}

Figure~\ref{Fig14} shows the hospitalization costs associated with the four policies. We can see that, in relative terms, the reduction of hospitalisation costs reflects quite closely the reduction of unmet care need.  In our model, the probability of being hospitalized depends positively on unmet care need. Therefore, the reduction of unmet care need leads directly to a reduction of hospitalization. However, we can see that the reduction of hospitalization cost for policy P4 is smaller than what we would expect looking at the unmet care need associated with this policy. If we consider that the probability of being hospitalized depends also on the level of social care need, we can see that the figure confirms the analysis of the previous figure: the child care policies are more effective than social care policies in preventing people from developing more serious health conditions and, therefore, will be more effective in reducing the hospitalisation probability for similar reductions of total unmet care need.

\begin{figure*}[!ht]
\caption{{\bf Formal child care.}}
\includegraphics[width=0.9\linewidth]{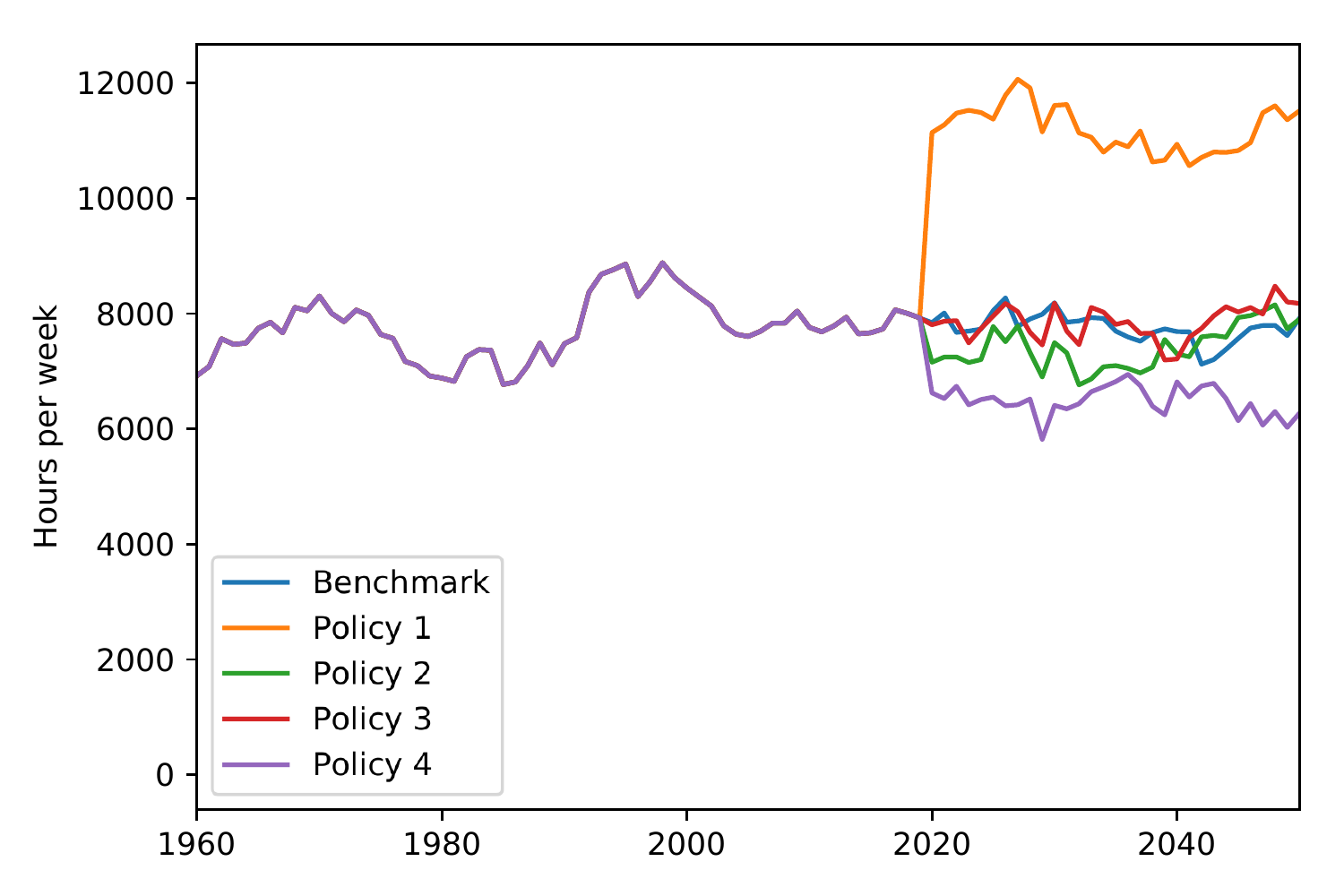}
\label{Fig15}
\end{figure*}

\begin{figure*}[!ht]
\caption{{\bf Share informal child care.}}
\includegraphics[width=0.9\linewidth]{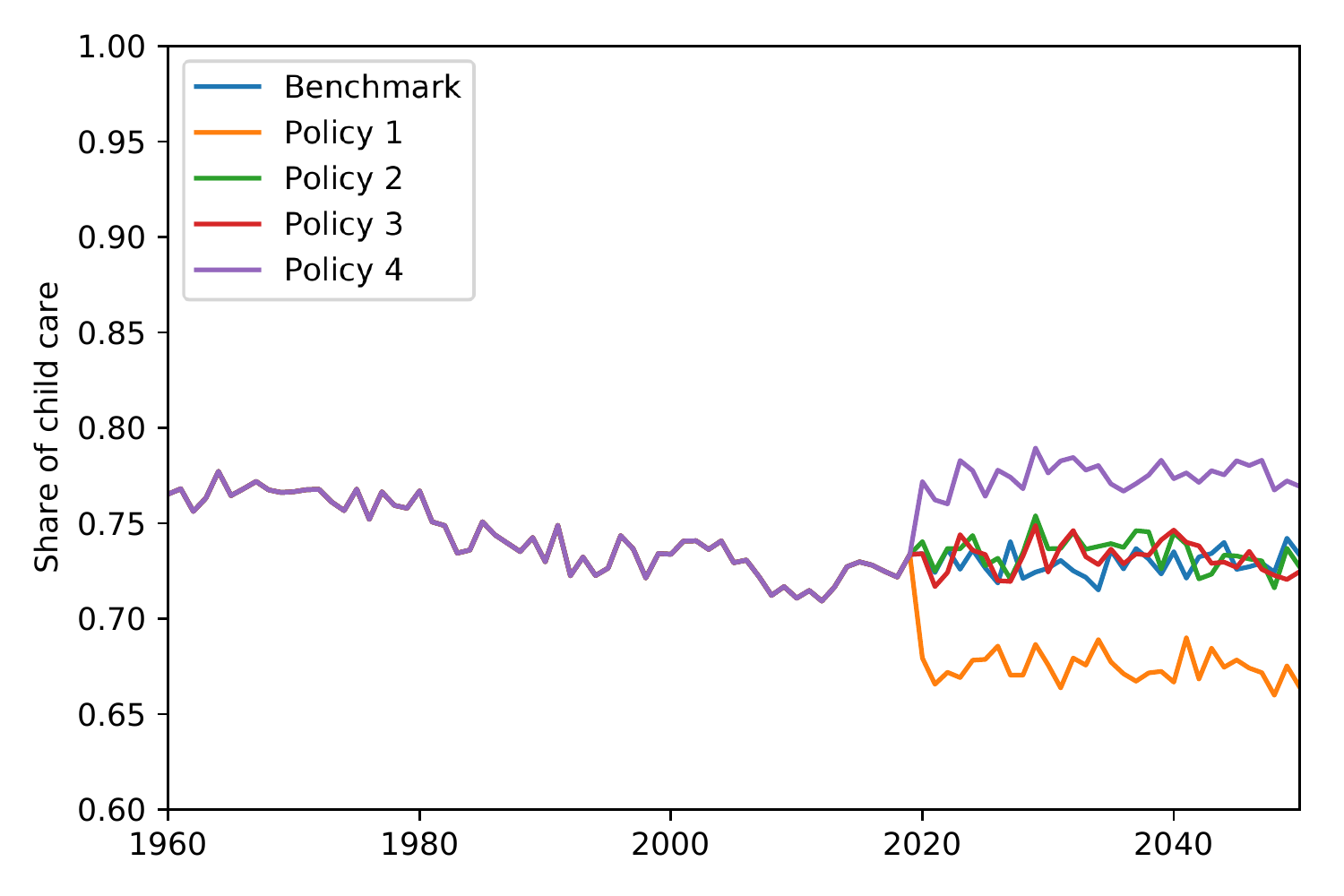}
\label{Fig16}
\end{figure*}

The next two figures show the opposite effect of policies P1 and P4 on formal and informal child care (while the previous figures showed the total effect over the period 2020-2050, these figures show the effect through time, 2020 being the year of policy implementation). Policy P1, making child care cheaper, generates an increase in the amount of formal child care, as we can see in Figure~\ref{Fig15}. On the other hand, policy P4 generates a decrease in formal child care. This spillover effect is due to the fact that when formal social care becomes cheaper, they can allocate time to child care rather than social care, and therefore the amount of formal child care decreases. This effect is shown also in Figure~\ref{Fig16}, where we can see that policy P1 reduces the share of child care represented by informal care, while the opposite effect is generated by policy P4.

\begin{figure*}[!ht]
\caption{{\bf Informal social care: total period 2020-2050.}}
\includegraphics[width=0.9\linewidth]{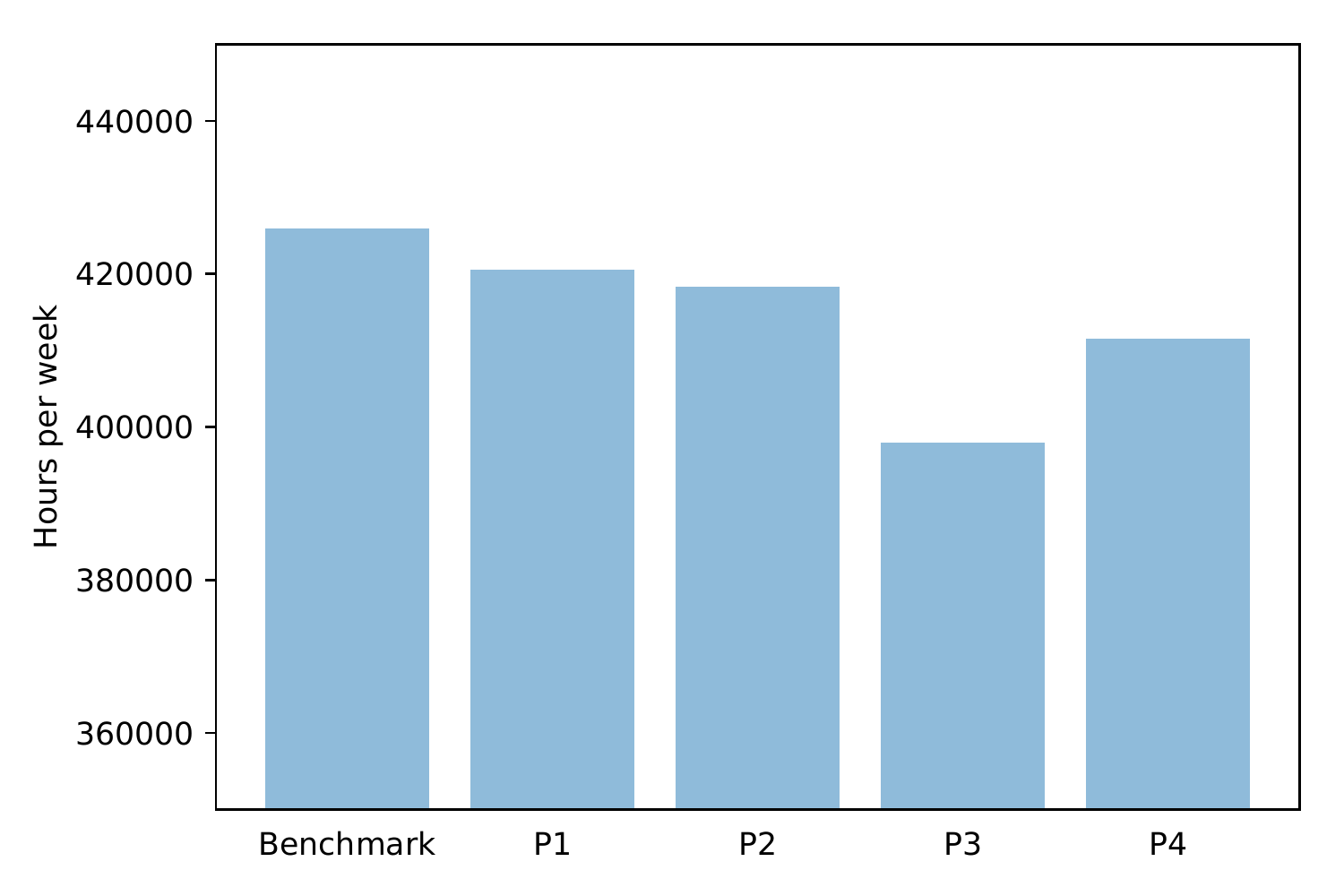}
\label{Fig17}
\end{figure*}

\begin{figure*}[!ht]
\caption{{\bf Formal social care: total period 2020-2050.}}
\includegraphics[width=0.9\linewidth]{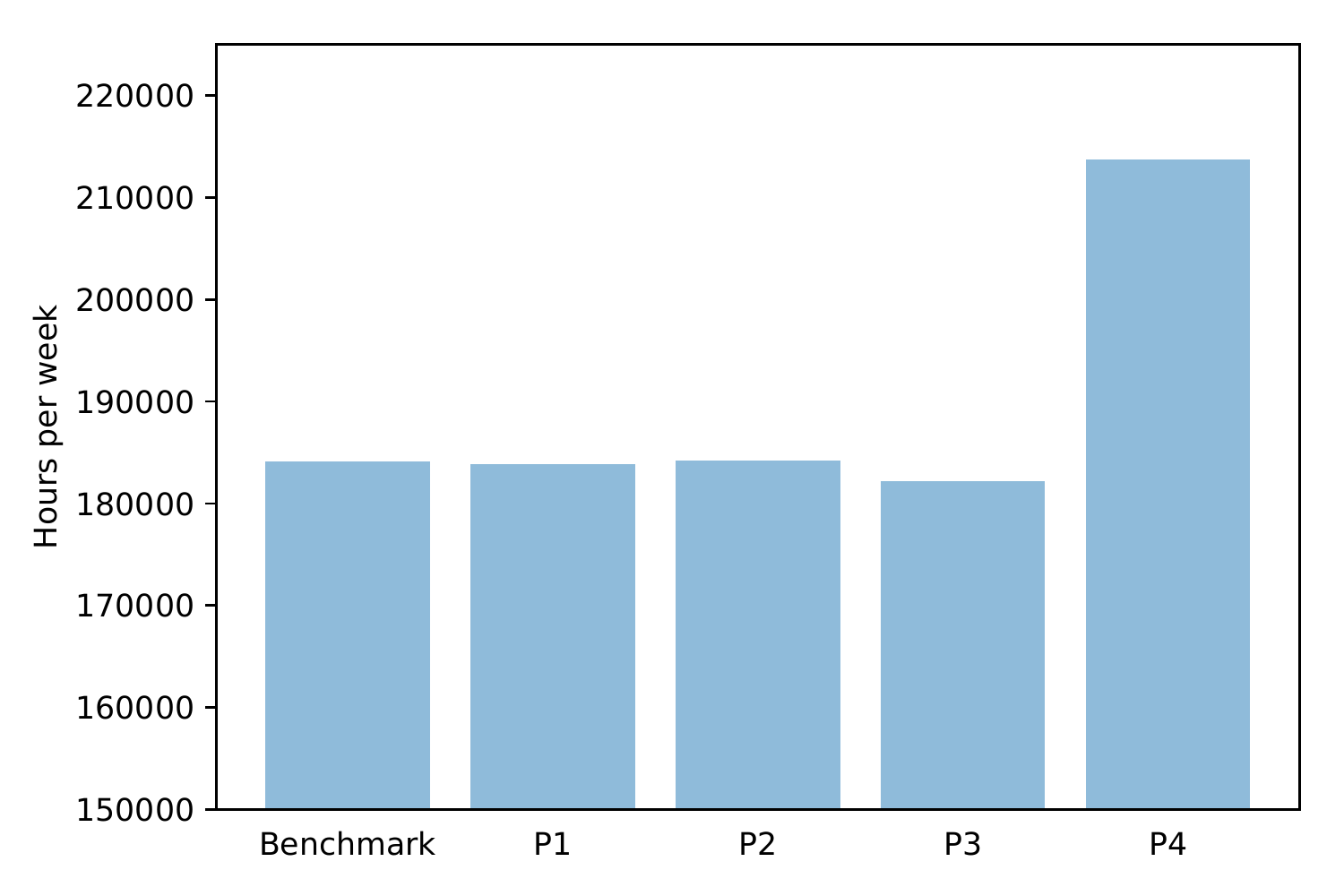}
\label{Fig18}
\end{figure*}

\begin{figure*}[!ht]
\caption{{\bf Public social care provided: total period 2020-2050.}}
\includegraphics[width=0.9\linewidth]{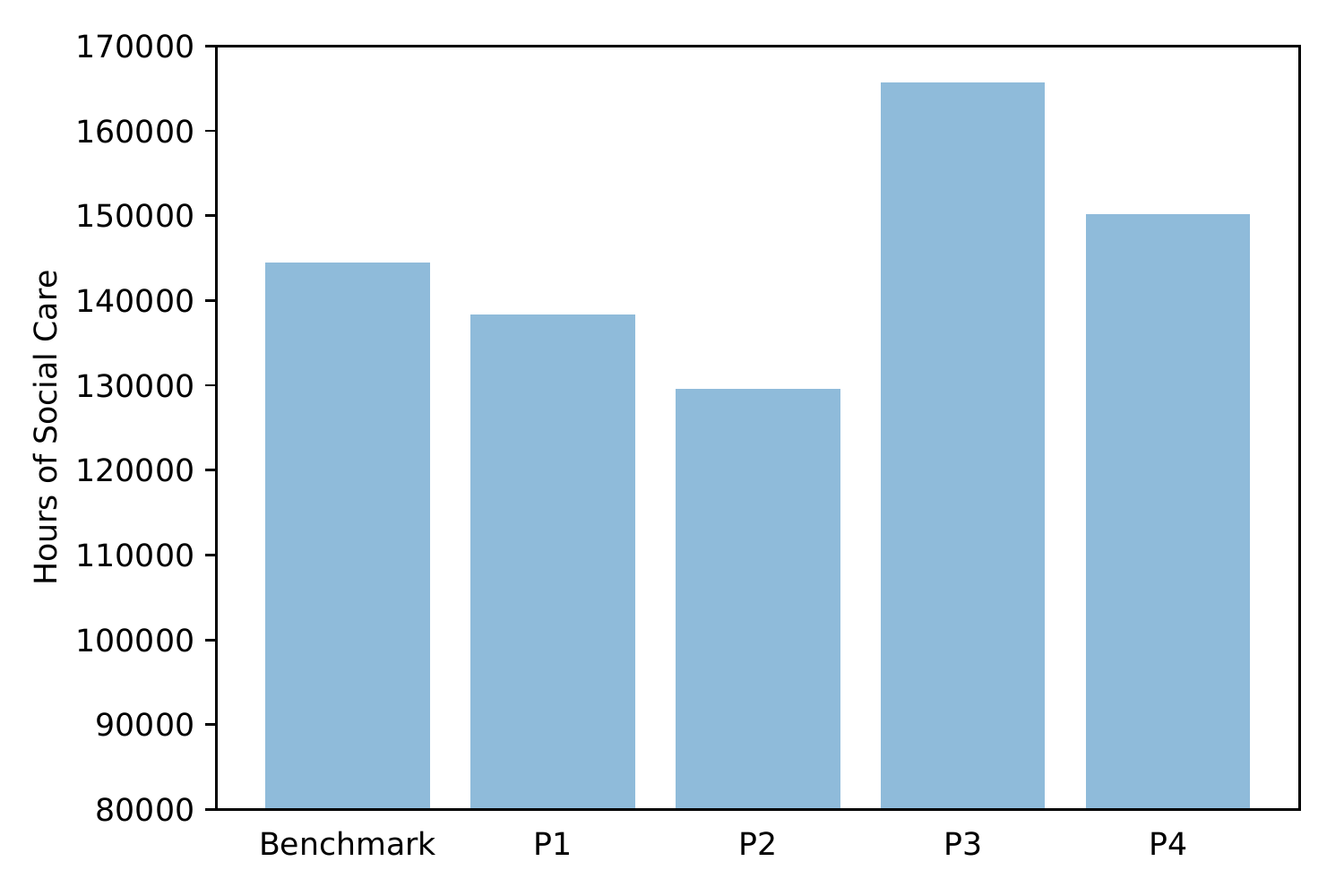}
\label{Fig19}
\end{figure*}

Figure~\ref{Fig17} shows the four policies' effect on informal social care provided in the period 2020-2050. We can see that policies P3 and P4 reduce the amount of informal care delivered. They cause this effect by increasing the other two forms of care: the formal care provided by the public sector in the case of policy P3 (as shown in Figure~\ref{Fig19}), and the privately-paid formal care in the case of policy P4, as shown in Figure~\ref{Fig18}.

Figure~\ref{Fig19} shows a comparison of public social care provided in the period 2020-2050.  The third policy (increasing eligibility for public care) is clearly the policy which most increases the public social care provided, as expected. More surprising is the reduction of public social care associated with the first two policies, which are meant to affect child care provision. This is another example of spillover effect generated, in this case, by the child care policies. This effect is due to the fact that the reduction of unmet care need shown in Figure~\ref{Fig11} reduces the share of people with the highest social care need level, and therefore the number of people eligible for the public social care provision. 

\begin{figure*}[!ht]
\caption{{\bf Off-work hours for care: total period 2020-2050.}}
\includegraphics[width=0.9\linewidth]{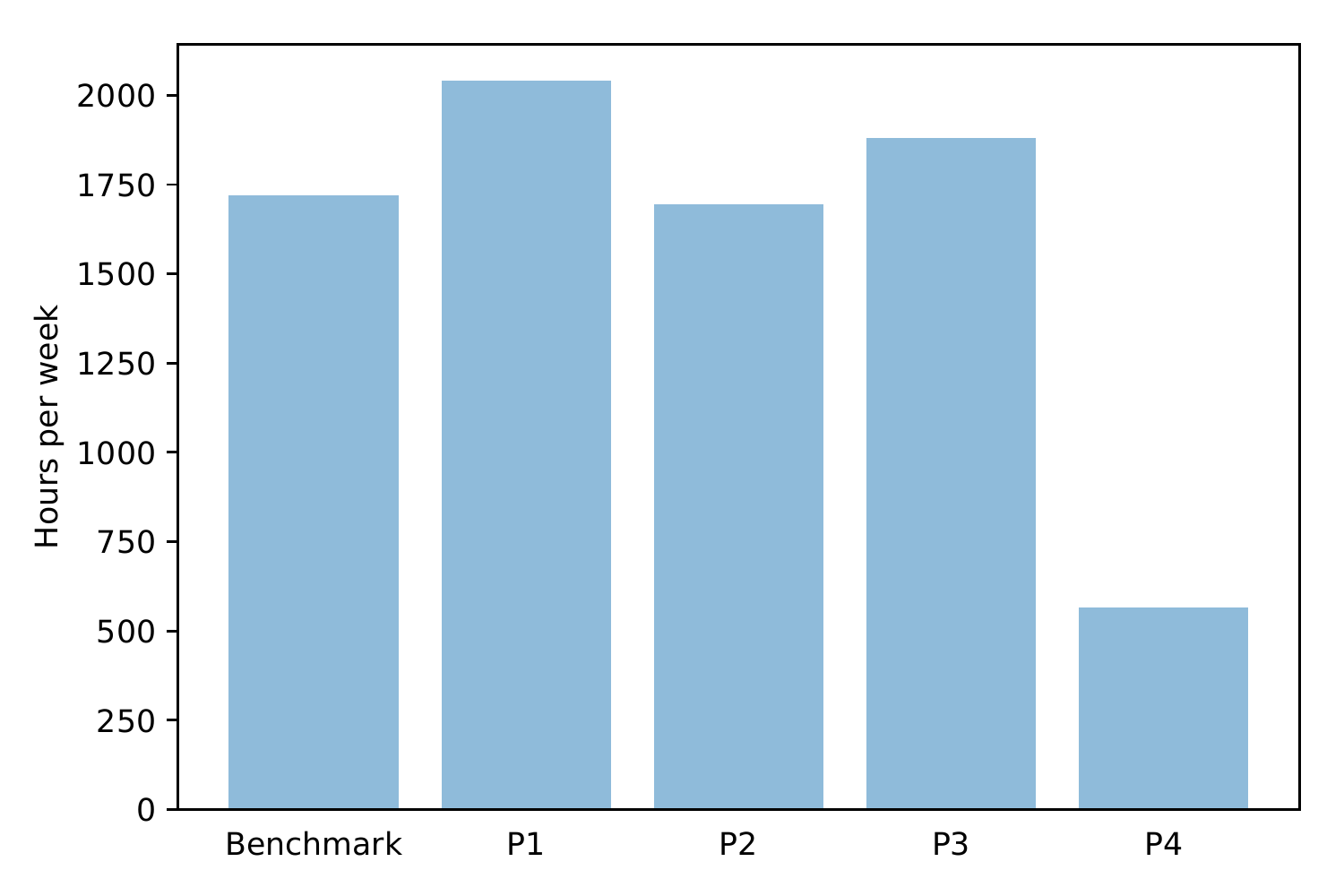}
\label{Fig20}
\end{figure*}

Finally, Figure~\ref{Fig20} shows the effect of the policies considered on the hours taken off work to provide care. We can see that policy P4 reduces the hours taken off work, due to the reduction of the cost of social care resulting from that policy. In fact, when the hourly cost of social care decreases, more people will prefer to work and pay for formal care, rather than taking hours off work to provide informal care.  The policy P1 also seems to increase the hours taken off work. This is due to the fact that P1 allows families to save income, so that they will have more financial resources to provide for social care. However, for families in the lower SESs (i.e. with hourly wages which are below the hourly price of social care), the increased financial availability will allow the working family members to take more hours off work to provide for social care.

\section*{Conclusion}

Here we have presented a detailed ABM of child and social care in a simulated UK population. We conducted a series of policy experiments to illustrate how the model can be used to compare the effects of different policies, allowing the policy-maker to investigate possible spillover effects and unintended consequences of polices before implementation in the real world.  We propose that this ABM can be a valuable tool for policy development and evaluation, as it explicitly models the complex interactions between child care and social care provision, and the negotiations that happen within families as they decide whether and how to allocate their time and money to care provision.  As a consequence of this detailed modelling of care decision-making and the effects of macro-level social policies, we can provide more sophisticated evaluations of policies that illuminate both their impact on government finances and their social and economic ramifications.

One of the strengths of ABM is that it allows us to discover possible spillover effects that may arise due to a policy change, given its ability to explicitly model complex interactions between policies.  Rolling out new policies is a slow, expensive and challenging process, as are revamping or retracting those policies in the event of unintended consequences; being able to test policies in simulation and uncover any spillover effects in advance could help policy-makers to avoid significant and costly problems after implementation.  In this paper we were able to demonstrate that the interrelationship of child care and social care can lead to unexpected consequences when one or the other of these processes is targeted by a policy intervention.  This in turn suggests that the best path toward reducing unment care need may come from surprising directions, and that policy-making in this area would thus benefit from this comprehensive modelling of the interactions between various kinds of care within families. 

In future work, we will continue to refine this modelling framework to allow users to more easily construct policy scenarios for evaluation.  We will enable the model to be adapted to other countries' social care systems by replacing the map and the mortality/fertility rates, and by implementing new social care policies.  We will also collaborate with social care policy experts and researchers to more accurately parameterise model processes.  Once the simulation framework is fully mature, we will generate more detailed and robust analyses of proposed real-world policy interventions directed at child and social care, both within the UK and elsewhere.

\section*{Acknowledgments}
Umberto Gostoli and Eric Silverman are part of the Complexity in Health Improvement Programme supported by the Medical Research Council (MC\_UU\_12017/14) and the Chief Scientist Office (SPHSU14).

This work was also supported by UK Prevention Research Partnership MR/S037594/1, which is funded by the British Heart Foundation, Cancer Research UK, Chief Scientist Office of the Scottish Government Health and Social Care Directorates, Engineering and Physical Sciences Research Council, Economic and Social Research Council, Health and Social Care Research and Development Division (Welsh Government), Medical Research Council, National Institute for Health Research, Natural Environment Research Council, Public Health Agency (Northern Ireland), The Health Foundation and Wellcome.

\bibliography{biblio}

\end{document}